\documentclass[twocolumn,epjc3]{svjour3}  
\RequirePackage{graphicx,subfig,multirow,ulem}
\RequirePackage{mathptmx} 
\RequirePackage[numbers,sort&compress]{natbib}
\RequirePackage[colorlinks,citecolor=blue,urlcolor=blue,linkcolor=blue]{hyperref}
\smartqed  
\usepackage{booktabs}
\usepackage{graphics}
\usepackage{comment}
\usepackage{lineno}
\usepackage[percent]{overpic}
\usepackage{amsmath}
\usepackage{amssymb}
\usepackage{wasysym}
\usepackage{rotating}
\usepackage{multirow}
\usepackage{overpic}

\RequirePackage{latexsym}
\RequirePackage[numbers,sort&compress]{natbib}
\RequirePackage[colorlinks,citecolor=blue,urlcolor=blue,linkcolor=blue]{hyperref}
\journalname{Eur. Phys. J. C}

\def\tectn{$^{130}$Te}

\def\pbdd{$^{210}$Pb }
\def\pbddn{$^{210}$Pb}

\def\udt{$^{238}$U }

\def\thdt{$^{232}$Th }

\def\tldd{$^{210}$Tl }
\def\tld{$^{208}$Tl }
\def\tldn{$^{208}$Tl}
\def\bidd{$^{212}$Bi }
\def\bidq{$^{214}$Bi }

\def\BBz{$0\nu\beta\beta$~}
\def\BBzn{$0\nu\beta\beta$}
\def\BBd{$2\nu\beta\beta$~}
\def\BBdn{$2\nu\beta\beta$}
\def\BB{$\beta\beta$~}

\def\Tz{$T^{0\nu}_{1/2}$~}
\def\Tzn{$T^{0\nu}_{1/2}$}
\def\Td{$T^{2\nu}_{1/2}$~}
\def\Tdn{$T^{2\nu}_{1/2}$}

\def\ca{$\sim$}

\def\teod{TeO$_2$~}
\def\teodn{TeO$_2$}
\def\be{\begin{equation}}
\def\ee{\end{equation}}

\def\ciccio{5$\times$5$\times$5 cm$^3$ }
\def\ciccion{5$\times$5$\times$5 cm$^3$}

\def\dmud{$\Delta m^{2}_{12}~$}

\def\dmdtn{$|\Delta m^{2}_{23}|$}

\def\tungstato{CdWO$_4$}
\def\zincato{ZnMoO$_4$}

\begin{document}

\title{Exploring the Neutrinoless Double Beta Decay in the Inverted Neutrino Hierarchy with Bolometric Detectors}

\author{D.~R.~Artusa\thanksref{USC,LNGS}
\and
F.~T.~Avignone~III\thanksref{USC}
\and
O.~Azzolini\thanksref{INFNLegnaro}
\and
M.~Balata\thanksref{LNGS}
\and
T.~I.~Banks\thanksref{BerkeleyPhys,LBNLNucSci,LNGS}
\and
G.~Bari\thanksref{INFNBologna}
\and
J.~Beeman\thanksref{LBNLMatSci}
\and
F.~Bellini\thanksref{Roma,INFNRoma}
\and
A.~Bersani\thanksref{INFNGenova}
\and
M.~Biassoni\thanksref{Milano,INFNMiB}
% ...
}

\thankstext{e1}{e-mail: cuore-spokeperson@lngs.infn.it}
\thankstext{d1}{Deceased}

\institute{
Department of Physics and Astronomy, University of South Carolina, Columbia, SC 29208 - USA\label{USC}
	\and
	INFN - Laboratori Nazionali del Gran Sasso, Assergi (L'Aquila) I-67010 - Italy\label{LNGS}
\and
INFN - Laboratori Nazionali di Legnaro, Legnaro (Padova) I-35020 - Italy\label{INFNLegnaro}
\and
Department of Physics, University of California, Berkeley, CA 94720 - USA\label{BerkeleyPhys}
\and
Nuclear Science Division, Lawrence Berkeley National Laboratory, Berkeley, CA 94720 - USA\label{LBNLNucSci}
\and
INFN - Sezione di Bologna, Bologna I-40127 - Italy\label{INFNBologna}
\and
Materials Science Division, Lawrence Berkeley National Laboratory, Berkeley, CA 94720 - USA\label{LBNLMatSci}
\and
Dipartimento di Fisica, Sapienza Universit\`a di Roma, Roma I-00185 - Italy\label{Roma}
\and
INFN - Sezione di Roma, Roma I-00185 - Italy\label{INFNRoma}
\and
INFN - Sezione di Genova, Genova I-16146 - Italy\label{INFNGenova}
\and
Dipartimento di Fisica, Universit\`a di Milano-Bicocca, Milano I-20126 - Italy\label{Milano}
\and
INFN - Sezione di Milano Bicocca, Milano I-20126 - Italy\label{INFNMiB}
% ...
}

\maketitle

\begin{abstract}
Neutrinoless double beta decay (\BBzn) is one of the most sensitive probes for physics beyond the Standard Model, providing
unique information on the nature of neutrinos. In this paper we review the status and outlook for bolometric \BBz decay searches. We summarize recent advances in background suppression demonstrated using bolo\-meters with simultaneous readout of heat and light signals. We simulate several configurations of a future
CUORE-like bolometer array which would utilize these improvements and present the sensitivity reach of a hypothetical next-gene\-ration bolometric \BBz experiment. We demonstrate that a bolometric experiment with the isotope mass of about 1 ton is capable of reaching the sensitivity to the effective Majorana neutrino mass ($|m_{ee}|$) of order 10-20 meV, thus completely exploring the so-called inverted neutrino mass hierarchy region. We highlight the main challenges and identify priorities for an R\&D program addressing them.
\end{abstract}

%
\section{Introduction}\label{sec:intro}
%
Neutrino oscillation experiments have provided compelling experimental evidence that neutrinos are massive and exhibit flavor mixing, but the absolute mass scale and the quantum nature of these particles (that is, if they are Dirac or Majorana fermions) remain unknown.  

The square mass differences \dmud and \dmdtn\ measured by neutrino oscillation experiments leave open three different possibilities for the ordering of the neutrino masses: normal hierarchy (NH), with $m_1<m_2 \ll m_3$, inverted hierarchy (IH), with $m_3 \ll m_1<m_2$ and  degenerate hierarchy (DH), with $m_1 \simeq m_2 \simeq m_3$.
However, oscillation experiments are not able to measure two fundamental properties of the neutrino: its nature (i.e. its quantum field structure) and its absolute mass. 
The most promising known way to investigate the Dirac-or-Majorana types of neutrino is neutrinoless double beta decay (\BBzn). If the neutrino is indeed found to be Majorana, then one can simultaneously achieve constraints on the absolute mass scale. Observation of \BBz process would demonstrate unambiguously that the lepton number is not strictly conserved. Such discovery would lend corroborating evidence to the leptogenesis hypothesis as a mechanism for generating matter-antimatter asymmetry in the Universe. 

Currently operating \BBz experiments probe the effective Majorana masses in the DH region. Several ambitious projects have been proposed with sensitivity in the IH region. Such experiments are very challenging, as they require significant detector masses and very low background levels. In this paper we discuss a ton-scale bolometric \BBz experiment based on simultaneous readout of both heat and light signals. This powerful technique offers superior background suppression, and the ability to investigate different \BBz candidate nuclei in large detector arrays. We present results of simulations focusing on the background levels that can be reached in specific experimental configurations and discuss their discovery potential and the ultimate sensitivity.

%
\section{Neutrinoless Double Beta Decay}\label{sec:dbd}
%
The transition in which an even-even nucleus (A,~Z) decays into its (A,~Z+2) isobar can be observed for isotopes whose single beta decay is forbidden. 
In the Standard Model (SM) this process is allowed  with the simultaneous emission of 2 electrons and 2 anti-neutrinos (\BBdn),  
and it has been  observed experimentally in more than ten isotopes with half-lives of the order of $10^{18}-10^{21}$\,y~\cite{Beringer:2012pdg}.

There are various hypothesized mechanisms for \BBzn~\cite{Bilenky:2012qi,Strumia:2006db}, all of them requiring physics beyond the Standard Model (SM). 
In particular, since the discovery of neutrino mass, the mechanism of virtual exchange of massive Majorana neutrinos has received increased attention.  This mechanism relates the \BBz decay half-life to important neutrino physics parameters which, in turn, help formulate detection strategies. 
The \BBz decay rate is proportional to the square of the so-called effective Majorana mass $|m_{ee}|$: 

\begin{equation}
\frac{1}{T_{1/2}^{0\nu }} = \frac{|m_{ee}|^2}{m_e^2} F^{0\nu}_N = \frac{|m_{ee}|^2}{m_e^2} G^{0\nu}\vert M^{0\nu} \vert ^2.
\label{eq:vita} 
\end{equation}

\noindent The quantity $|m_{ee}|$ is defined in terms of the three neutrino masses and of the elements of the Pontecorvo-Maki-Nakaga\-wa-Sakata (PMNS) matrix~\cite{Maki:1962mu}, as follows:

\begin{equation}
|m_{ee}| = \left |\sum_k U_{ek}^2 m_k \right |~.
\label{eq:meedef} 
\end{equation}

In Eq.~(\ref{eq:vita}), \Tz is the decay half life, $m_e$ is the electron mass, $G^{0\nu}$ is the two-body phase-space factor, 
and  $M^{0\nu}$ is the \BBz nuclear matrix element (NME). The product $F^{0\nu}_N = G^{0\nu} |M^{0\nu}|^2$ is referred to as nuclear  factor of merit.
While $G^{0\nu}$ can be calculated with reasonable accuracy, the NME value is strongly dependent on the nuclear model used for its evaluation. 
This problem, which is discussed in more detail in Section~\ref{sec:NME}, adds considerable uncertainties in the calculation of $|m_{ee}|$  from experimental measurements or limits on the half-life. 

\begin{figure}[t!]
 \begin{center}
 \includegraphics[width=0.48\textwidth,clip=true]{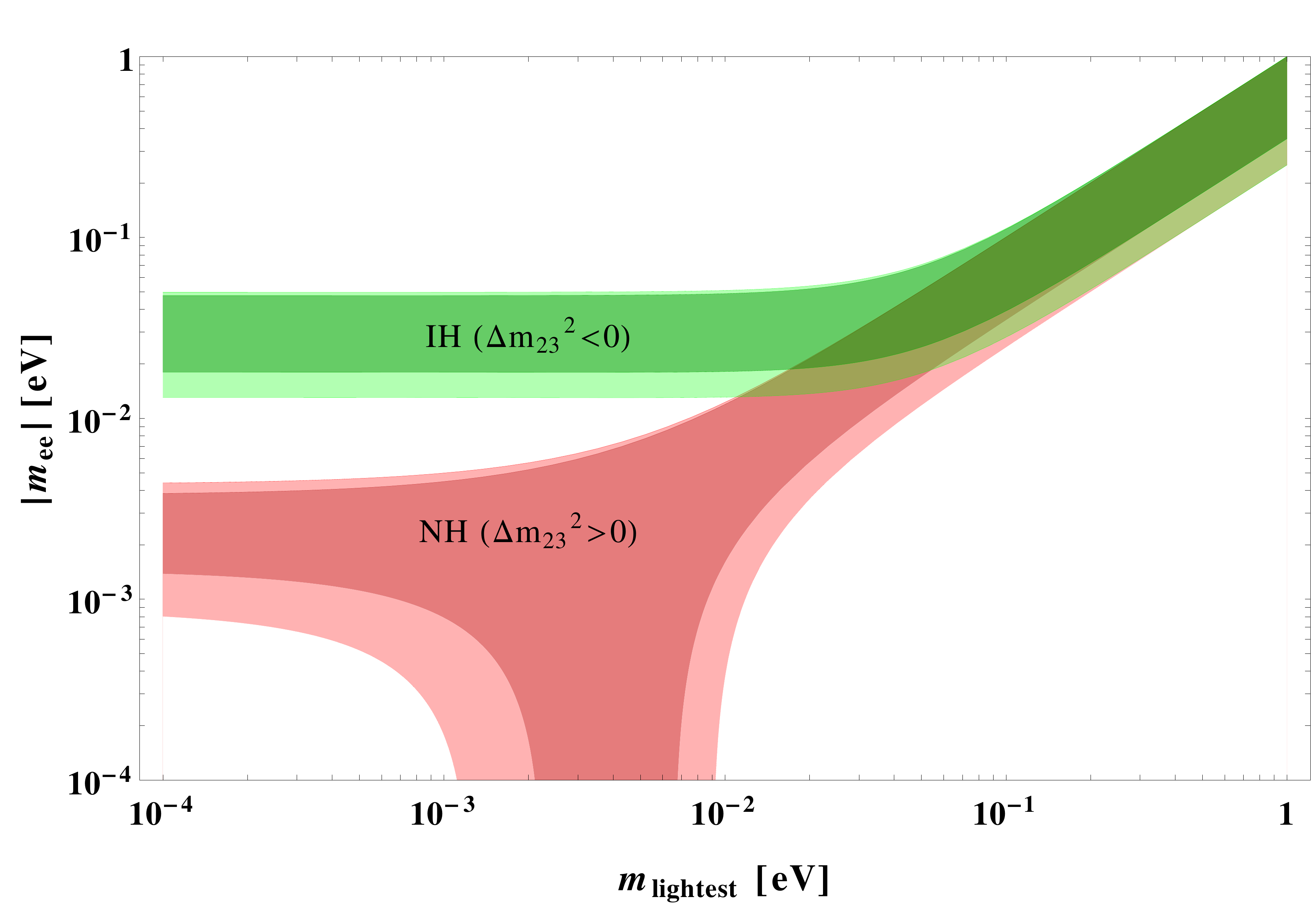}
 \end{center}
\caption{Majorana neutrino mass as a function of the lightest neutrino mass in the normal hierarchy ($\Delta m^2_{23}>0$) and in the inverted hierarchy ($\Delta m^2_{23}<0$) scenarios. The shaded areas correspond to the 3$\sigma$ regions due to error propagation. Figure from~\cite{Vissani2014aa}.}
\label{fig:mbb_vs_ml} 
\end{figure}

Given the present experimental results on the parameters governing neutrino oscillations, the expected allowed ranges for $|m_{ee}|$ as a function of the lightest neutrino mass are depicted in Fig.~\ref{fig:mbb_vs_ml}. 
For the inverted hierarchy scenario, the range of the possible values for the effective Majorana mass is~\cite{Bilenky:2012qi,Strumia:2006db,Vissani2014aa}:
\begin{equation}
10 \lesssim |m_{ee}| \lesssim 50\; \rm{meV}\;.
\label{eq:rangeIH} 
\end{equation}
One of the priorities in neutrino physics is the experimental determination of the neutrino mass hierarchy. Neutrino oscillation experiments (accelerator-driven long-baseline experiments, measurements at the reactors, as well as high-statistics studies of atmospheric neutrinos) and cosmological constraints on the sum of neutrino masses have the potential to resolve the ordering of neutrino mass states in the next decade or two~\cite{Cahn:2013aa}. Thus, searches for \BBz decays with the sensitivity to completely explore the inverted hierarchy region are both relevant and timely. Given the spread in the values of the nuclear matrix elements (Section~\ref{sec:NME}), achieving this goal requires that the next-generation experiments aim at a sensitivity well below $|m_{ee}|<20$~meV. 

%
\subsection{Experimental sensitivity} \label{sec:expsensitivity}
%
Experimentally, neutrinoless double beta decay searches rely on the measurement of the two emitted electrons. 
In the so-called homogeneous approach (source = detector, Sec.~\ref{sec:sart}) one detects the two electrons in the same detector volume.  The summed kinetic energy of the two electrons and nuclear recoil is equal to the Q-value of the \BBz transition, which is energetically dominated by the electrons. The signal would thus appear as a peak at the energy of the Q-value. Existing limits constrain \BBz decay, if it occurs at all, to be extremely rare. 

Observing potential \BBz counts is hindered by background events in the signal region of interest (ROI). If a \BBz peak is observed in the measured energy spectrum, the half-life can be evaluated as:
\begin{equation}
T^{0\nu}_{1/2} = \ln 2\, T  \epsilon \,N_{\beta\beta}  / N_{\rm{peak}}
\label{eq:t0nu}
\end{equation}

\noindent where $T$ is the measuring time, $\epsilon$ is the detection efficiency, $N_{\beta\beta}$ is the number of \BB source nuclei under observation, and $N_{\rm{peak}}$ is the number of observed \BBz decays.

If no peak is detected, the sensitivity of a given \BBz experiment is usually expressed in term of the detector sensitivity $\widehat{T^{0\nu}_{1/2}}$ at $n_{\sigma}$ (we use hatted quantities to indicate sensitivities instead of true values), defined as the half-life corresponding to the signal that could be emulated by a background fluctuation of a chosen significance level, expressed in numbers of Gaussian standard deviations ($n_{\sigma}$), in the limit of large background in the ROI:

\begin{equation}
\widehat{T^{0\nu}_{1/2}} (n_\sigma) = \frac{\ln(2)}{n_\sigma} \frac{N_A ~ a ~\eta ~ \epsilon}{W} f(\Delta E) 
\sqrt{ \frac{ M ~ T }{B ~ \Delta E}}  ~,
\label{eq:sensitivity}
\end{equation}

\noindent where $\eta$ is the stoichiometric coefficient of the \BB candidate; $a$ is the \BB candidate isotopic abundance; $N_A$ is Avogadro's number; $W$ is the molecular weight of the active mass; $B$ is the background per unit of mass, time and energy; $M$ is the detector mass; $T$ is the live time; $\Delta$E is the ROI energy window (typically FWHM energy resolution); $\epsilon$ is the detector efficiency; and $f(\Delta E)$ is the fraction of signal events that fall in an energy window $\Delta E$ around the Q-value.

Finally, $\widehat{T^{0\nu}_{1/2}}$ from Eq.~(\ref{eq:sensitivity}), along with $F^{0\nu}_N$ from Eq.~(\ref{eq:vita}), is translated into an effective Majorana mass sensitivity\footnote{When using Eq.~\ref{eq:mee} to convert the sensitivity $\widehat{T^{0\nu}_{1/2}}$ into a Majorana mass range (for example in Table~\ref{tab:resultsIHE} and \ref{tab:results}) we apply the correct error propagation, according to~\cite{Beringer:2012pdg}.}:
\begin{equation}
|m_{ee}| \propto \frac{m_e}{(\widehat{T^{0\nu}_{1/2}} F^{0\nu}_N)^{1/2}}
\label{eq:mee}    
\end{equation}
which, again referring to Eq.~(\ref{eq:sensitivity}), highlights the slow dependence (fourth root) of $|m_{ee}|$ on the experimental parameters $M$, $T$, $B$, and $\Delta E$ in the limit of large background.

Eq.~(\ref{eq:sensitivity}) holds if the number of background counts is large enough that its distribution can be considered to be Gaussian. In the so-called zero background limit, when the expected number of background counts is small, one should use Poisson statistics and the corresponding formula for the sensitivity at a given credibility level (c.l.) is:

\begin{equation}
\widehat{T^{0\nu}_{1/2}} {\rm{(c.l.)}} = -\frac{\ln(2)}{\ln(1-\frac{\rm{c.l.}}{100})} \frac{N_A ~ a ~\eta ~ \epsilon}{W} \cdot M \cdot T \cdot f(\Delta E) ~ .
\label{eq:sensitivityzero}
\end{equation}

%
 \subsection{Nuclear Matrix Element}\label{sec:NME}
%
It is clear from Eq.~(\ref{eq:vita}) that the evaluation of the nuclear matrix element is needed in order to either: 1) extract the value of $|m_{ee}|$ from the experimentally measured \BBz decay rate or 2) convert an upper limit on the \BBz decay rate to an upper limit on $|m_{ee}|$. Any uncertainty in the calculated values of $M^{0\nu}$ will correspond to a significant uncertainty on $|m_{ee}|$. Moreover, knowledge of the NME is required both when planning new experiments and also when comparing the results of experiments that use different nuclei. 

\begin{figure}
 \begin{center}
 \includegraphics[width=0.5\textwidth,clip=true]{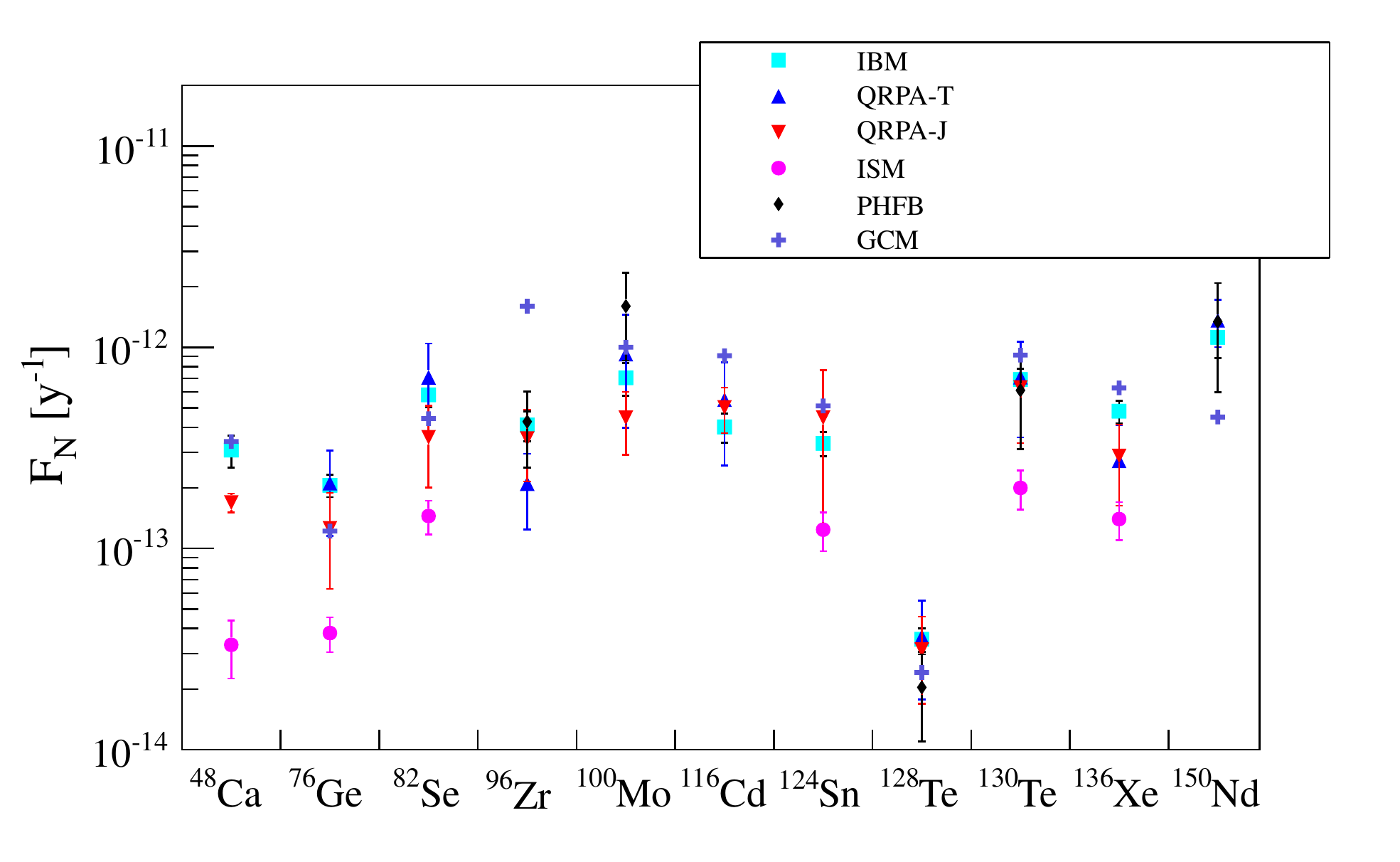}
 \end{center}
\caption{Nuclear factor of merit values calculated for different most recent theoretical models and for many different \BBz candidates. The bars represent the spread of the models.}
\label{fig:FN} 
\end{figure}

The calculation of $M^{0\nu}$ requires an accurate nuclear mo\-del. Since all \BBz candidate nuclei are relatively heavy, the corresponding many body problem cannot be solved without approximations. Fig.~\ref{fig:FN} presents a summary of calculated $F^{0\nu}_N$ results for many different \BBz candidates based on recent publications. For each isotope, the available results from the following nuclear models have been considered: Quasi Particle Random Phase Approximation (QRPA)~\cite{QRPAT:2012,QRPAT150Nd:2011,QRPAJ:2012}, Interacting Shell Model (ISM)~\cite{Menendez:2008jp}, microscopic Inte\-racting Boson Model (IBM-2)~\cite{IBM:2013}, Projected Hartree-Fock-Bogoliubov model (PHFB)~\cite{PHFB:2010} and Generating Coordinate Method (GCM)~\cite{GCM:2010}. Values for $G^{0\nu}$ were taken from~\cite{IachelloPSF:2012}.

\subsection{State of the Art}\label{sec:sart}

A variety of detection techniques are used in \BBz experiments. In the so-called homogeneous experiments the active volume of the detector contains the double beta decaying isotope (source = detector) and, in case a decay occurs, the sum energy of the two emitted electrons is detected.
Homogeneous experiments can be based on solid (bolometers, semiconductors, or scintillators), liquid (TPC: Time Projection Chamber or scintillators), and gaseous (TPC) devices. In tracking experiments the double beta emitting isotope is contained in thin foils surrounded by tracking detectors. With this approach the emitted electrons can be identified and tracked separately. In some cases, like the proposed NEXT experiment~\cite{NEXT}, a homogeneous detector can also be operated as a tracking device.

Apart from one unverified claim\footnote{\label{klapdor-claim} After the publication of the last official results of the thirteen-year-long Heidelberg-Moscow $^{76}$Ge experiment \cite{Klapdor:2000sn}, a part of the collaboration reanalyzed the full data set of 71.7 kg$\times$y exposure and claimed a 4.2\,$\sigma$  
observed signal for \BBz in $^{76}$Ge, with $T^{0\nu}_{1/2}=1.19\times 10^{25}$y \cite{Klapdor:2004wj}. More recently \cite{Klapdor:2006ff}, continued refinement of their methods resulted in a published confidence level of 6.4\,$\sigma$ corresponding to a decay half-life of  $T^{0\nu}_{1/2}=2.23^{+0.44}_{-0.31}\times 10^{25}$y and a $|m_{ee}|$ value between 0.19 and 0.59\,eV (according to the $F^{0\nu}_N$ values of Fig.~\ref{fig:FN}). Since its first appearance, the claim has been strongly criticized by the double beta decay community because of the assumptions made in the background evaluation procedure \cite{Vissani3,Vissani4}. Nevertheless, regardless of its credibility, all future double beta decay experiments will necessarily have to compare with this result.},
no experimental evidence for neutrinoless double beta decay has been observed so far. Experimental half-life lower limits have been obtained for several isotopes:  $^{76}$Ge~\cite{Klapdor:2000sn,Aalseth:2002rf,Agostini:2013mzu}, $^{82}$Se~\cite{Arnold:2006sd,Barabash:2010bd}, $^{100}$Mo~\cite{Arnold:2006sd,Barabash:2010bd}, $^{130}$Te~\cite{Andreotti:2010vj}, and $^{136}$Xe~\cite{KamLANDZen:2012aa,Auger:2012ar}. 

Contemporary efforts are focused on so-called second generation experiments (CUORE~ \cite{ACryo}, SuperNEMO~\cite{Piquemal:2006cd}, nEXO~\cite{EXO}, NEXT~\cite{NEXT}, LUCIFER~\cite{Ferroni2010}, GERDA II~\cite{Abt:2004yk}, SNO+ \cite{SNO+}) with the goal of approaching the IH region at $|m_{ee}| \leq 50$\,meV. Some of these experiments have already started the first phase (EXO-200~\cite{Auger:2012ar}, KamLAND-Zen~\cite{KamLANDZen:2012aa}, GERDA-I~\cite{GERDA:2013}, MAJORANA Demonstrator~\cite{MJD:2013aa}) with reduced mass devoted to the investigation of the $^{76}$Ge positive claim\footnote{In July 2013, the GERDA collaboration published its first result on the neutrinoless double beta decay of $^{76}$Ge~\cite{Agostini:2013mzu}. They observed no signal and set a lower limit $T^{0\nu}_{1/2} > 2.1\cdot10^{25}$ y (90\% C.L.). The combination with the results from previous $^{76}$Ge experiments yields $T^{0\nu}_{1/2} >  3.0\cdot10^{25}$ y (90\% C.L.).}, while considerable R\&D is devoted to new techniques which could contribute to the full exclusion of the IH mass region ($|m_{ee}| \leq 10$\,meV). 

%
\section{A strategy for the future: bolometers}\label{sec:bolo}
%
The realization of an experiment with a reasonable discovery potential down to the smallest $|m_{ee}|$ values is an incredible challenge in which detector technology plays a key role. 

The most promising isotopes (see Table~\ref{tab:isotopi}) are those that show a high nuclear factor of merit\footnote{It has been pointed out in a recent work~\cite{Robertson:2013} that, due to an approximate inverse correlation between phase space and the square of the nuclear matrix element that emerges from existing calculations, no isotope is really favored or disfavored; all have qualitatively a similar decay rate per unit mass for any given value of the Majorana mass.}. However, even for these, the expected event rate from \BBz decay for a Majorana mass of 10 meV, is as low as 0.1 to 1 counts per year using a metric ton of source material (cnts/y/ton$_{\rm{iso}}$). Under these conditions, the exposure required to record even a few \BBz decays is of the order of 1 to 10 (y $\cdot$ ton$_{\rm{iso}}$). Moreover, to reach a reasonable signal-to-background ratio in the energy region $\Delta$E of approximately a FWHM around the Q-value, a background counting rate lower than 0.01 cnts/keV/y/ton$_{\rm{iso}}$ is needed for high resolution detectors ($\Delta$E \ca 10 keV) or lower than 0.001 cnts/keV/y/ton$_{\rm{iso}}$ for low resolution detectors \- ($\Delta$E \ca 100 keV). 
At these low rates the intrinsic background from \BBd decay can compromise the sensitivity of an experiment. For a given \Tzn, the ratio between the number of \BBd and \BBz decays in the energy region of interest for \BBz can be evaluated on the basis of the measured \Tdn. As shown in Table~\ref{tab:isotopi}, detectors with good energy resolution ($\Delta$E \ca 10 keV) have a \BBd to \BBz event rate ratio well below one. This is a very strong motivation for next generation experiments to exploit high resolution detectors if possible.

The impact of environmental $\gamma$ radiation on the background counting rate of a \BB experiment is an important consideration.  The most appealing isotopes to study are those with a Q-value above most of the natural $\gamma$-ray spectrum. The \tld line at 2.615 MeV is the most energetic $\gamma$ peak visible in environmental background spectra; above this line there are only extremely rare $\gamma$ rays (e.g. the very rare gamma lines of $^{214}$Bi) or gamma emission stimulated by n capture.  Therefore, isotopes with Q-values greater than 2.6 MeV are more desirable from this perspective.

Another important consideration is the natural isotopic abundance of the \BBz candidate of interest (the parameter $a$ in Eq.~(\ref{eq:sensitivity})) and the isotopic enrichment cost.  In Table~\ref{tab:isotopi}, the natural abundances of the most common \BBz nuclei are listed.
Enrichment of detectors is limited by cost and technical feasibility. For example, in the case of $^{48}$Ca and $^{150}$Nd, the standard high-volume enrichment techniques are difficult at present. However, it is feasible to enrich all the other isotopes listed in Table~\ref{tab:isotopi} (see Refs.~\cite{enr,enr-NEMO,enr-EXO,enr-Cd}). Among these isotopes, cost of enrichment for $^{130}$Te and $^{136}$Xe is relatively modest, but is more significant for others.

In summary, a detector's sensitivity to \BBz is directly related to the choice of the \BB isotope being studied, and to the detector technology being used.   

\begin{table*}
\caption{Properties of the most commonly studied \BBz candidates: Q-value, isotopic abundance, and \Td half-life (average values from~\cite{Barabash:2010ie}). R$_{0\nu}^{|m_{ee}|=10\rm{meV}}$ is the range of \BBz count rates expected in an energy window of a FHWM around the Q-value, for a Majorana mass of 10 meV, corresponding to the largest and smallest NME values among those of Fig.~\ref{fig:FN}. The ratio of signal counts in an energy window of a FHWM around the Q-value for a Gaussian distributed-signal is $f(\Delta E = {\rm FWHM}) = 0.76$. In the last column we report  the \BBd event rate around the Q-value, calculated by integrating the last 10 keV below the end point of the \BBd spectrum. All rates are expressed in counts per year per ton of \BBz emitting isotope.}
\begin{center}
\begin{tabular}{cccccc}
\hline\hline
Isotope & Q & a & \Td & R$_{0\nu}^{|m_{ee}|=10\rm{meV}}$ & R$_{2\nu}^{10\rm{keV}}$ \\ 
 &[keV]& [\%] &  10$^{19}$ [y] & [cnts/y/ton$_{\rm{iso}}$] & [cnts/y/ton$_{\rm{iso}}$] \\
\hline
\\
 $^{48}$Ca	& 4274  	& 0.2		&  4.4$^{+0.5}_{-0.4}$		& 0.06 - 0.9 	& 5 $\times$ 10$^{-6}$  	\\
 $^{76}$Ge	& 2039 	& 7.6 	& 160$^{+13}_{-10}$ 	 	& 0.05 - 0.5 	& 4 $\times$ 10$^{-6}$ 	\\
 $^{82}$Se 	& 2996 	& 8.7 	& 9.2$\pm$0.7	 	 	& 0.17 - 1.5 	& 8 $\times$ 10$^{-6}$	\\
 $^{96}$Zr		& 3348 	& 2.8 	& 2.3$\pm$0.2 			& 0.16 - 2.0    	& 2 $\times$ 10$^{-5}$ 	\\
$^{100}$Mo	& 3034 	& 9.6		& 0.71$\pm$0.04	 	& 0.35 - 2.9 	& 8 $\times$ 10$^{-5}$ 	\\
$^{116}$Cd	& 2814 	& 7.5 	& 2.85$\pm$0.15	 	& 0.27 - 0.9    	& 2 $\times$ 10$^{-5}$        \\
$^{130}$Te	& 2528 	& 34.2 	& 69$\pm$13	 		& 0.15 - 1.0    	& 2 $\times$ 10$^{-6}$ 	\\
$^{136}$Xe	& 2458 	& 8.9 	& 220$\pm$6 			& 0.1 - 0.6 	 	& 6 $\times$ 10$^{-7}$ 	\\
$^{150}$Nd	& 3368 	& 5.6 	& 0.82$\pm$0.9			& 0.36 - 1.7 	& 3 $\times$ 10$^{-5}$ 	\\
\\
\hline\hline
\end{tabular}
\end{center}
\label{tab:isotopi}
\end{table*}

Cryogenic bolometers \cite{bolo-review1,bolo-review2} have been used for many years as particle detectors to study rare events such as \BBz\, and dark matter. They are solid state devices, kept at a temperature of $\sim$10 mK, where a particle's kinetic energy is converted into lattice vibrations of the absorber material, generating a temperature rise ($\Delta T$). Such low temperature is necessary since the heat capacity of an insulating crystal varies as C $\propto T^3$. The $\Delta T$ (about 0.1 mK for 1 MeV of deposited energy in a 750 g crystal) is measured by a thermometer (e.g. a semiconductor thermistor) affixed to the surface of the absorber, that converts the temperature rise into an electric pulse.

The advantages of the bolometric technique in the field of \BBz have not been completely exploited. 
In particular, bolometers offer a wide choice of possible absorber materials while also being able to achieve an energy resolution competitive with that of Ge diodes (e.g. on the order of 5 keV FWHM at 3 MeV). The freedom in the choice of the absorber provides an opportunity for consistent methodology over a wide range of possible \BB isotope candidates without the limitations usually imposed by the experimental technique (e.g. as seen with the semiconductor detectors).

As a consequence, with bolometers it is possible to maximize the sensitivity in Eq.~(\ref{eq:sensitivity}) (or Eq.~(\ref{eq:sensitivityzero})) through the optimization of all its terms, something that is seldom achievable with other kinds of detectors. In particular:
\begin{enumerate}
\item{Bolometers can maximize efficiency and fiducial mass because they are solid state detectors; the large scale ($\sim 0.2$ ton$_{\rm{iso}}$) in a bolometric experiment is being tested by the CUORE~\cite{ACryo}  experiment currently under construction;}
\item{The energy resolution is among the best ever measured for massive solid state detectors, ensuring negligible background from \BBd spectrum tail;}
\item{The isotopic abundance can be maximized through enrichment (except for $^{48}$Ca and $^{150}$Nd with present technology);}
\item{As discussed later in this paper, the background can be minimized with a choice of the isotope and by employing active rejection techniques.}
\end{enumerate}

The CUORE experiment represents the most advanced stage in the use of bolometers for \BBz searches. CUORE will consist of an array of  \ca 1000 crystals for a total mass of \ca 1 ton of \teod and \ca 200 kg of \tectn. It is expected to be taking data at Laboratori Nazionali del Gran Sasso (LNGS) in 2015. The sensitivity of CUORE will depend on the background level, with the target near 50 meV~\cite{CUOREsensi}. In the future CUORE may lead the way toward a few-ton scale experiment capable of either exploring the entire IH region, or making a precision measurement of $T_{1/2}^{0\nu}$.

%
\subsection{Scintillating bolometers} \label{sec:scintbolo}
%
In the previous section we enumerated the many advantages of using bolometers for \BBz searches. Among those, we mentioned the possibility to actively reject radioactive background via particle identification; this is possible employing scintillating bolometers.

Scintillating bolometers, used already both for \BBzn~\cite{nsvecchioarticoloCaF2,Pirr06} and for dark matter~\cite{CRESST,ROSEBUD} searches, provide a mechanism to distinguish $\alpha$ interactions (which are part of the background only) from $\beta/\gamma$ interactions (which can be a part of both the background and signal). 
   
\begin{figure}
\begin{center}
\includegraphics[width=9cm,clip=true]{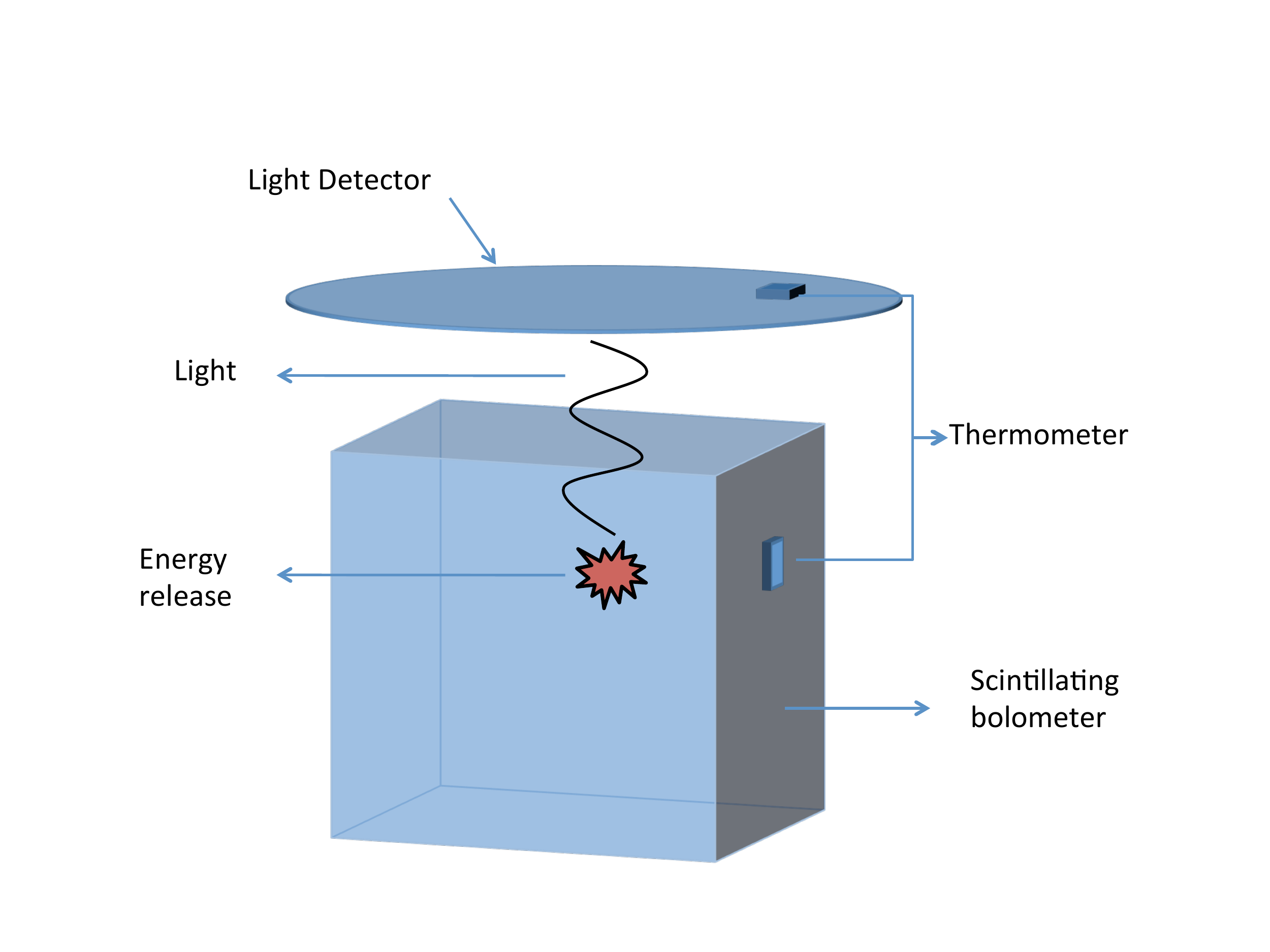}
\end{center}
\caption{Operating principle of a scintillating bolometer. The release of energy inside a scintillating crystal follows two channels: light production and thermal excitation.}
\label{fig:bol_scint}
\end{figure}

A scintillating bolometer functions by operating a scintillating crystal as a cryogenic bolometer (as described above) and coupling it to a light detector, as shown in Fig.~\ref{fig:bol_scint}.  Similar to other large-mass bolometers, the best energy resolution is achieved at extremely low temperatures (\ca10~mK). 

When a particle traverses the scintillating crystal and interacts with the lattice, a large fraction of the energy is transferred into the crystal as heat, raising the internal energy, thus inducing the already mentioned temperature rise.  A small fraction of the deposited energy produces scintillation light that propagates as photons out of the crystal.  These are then detected by a separate light detector facing the crystal. 
The light detectors used so far for scintillating bolometers are bolometers themselves and consist of germanium or silicon wafers, kept at the same temperature as the main bolometer. Scintillation photons deposit heat into the wafer and induce a temperature rise, which is then measured by a second thermistor.

The signals registered by the two thermistors are conventionally named heat (the one generated in the main bolometer) and light (the one induced in the light detector). Although they have the same nature (temperature rises), they originate from different processes.

An interesting feature of scintillating bolometers is that the ratio between the light and heat signals depends on the particle mass and charge. Indeed, while the thermal response of a bolometer has only a slight dependence on the particle type\footnote{This dependence is of the order of 7$\permil$ for \teod crystals~\cite{Beeman:2011yc} and about 10-20\% for scintillating crystals~\cite{Arnaboldi:2010jx,Beeman:2013znse}.}, the light emission from a scintillator changes significantly with ionization energy density. Particles like $\beta$s and $\gamma$s have similar light emission (referred to as the light yield, i.e.~the fraction of particle energy emitted in photons), which is typically different from the light emission induced by $\alpha$ particles or neutrons. 
Consequently, the coincident measurement of the heat and light signals allows particle discrimination. If the scintillating crystal contains a \BB candidate, the \BBz signal (i.e. the energy deposition produced by the two electrons emitted after the decay) can be distinguished from an $\alpha$ signal \cite{nsvecchioarticoloCaF2,Pirr06}, leaving only $\beta$s and $\gamma$s to give sizable contribution to the background. 

Scintillating bolometers containing Ca, Mo, Cd, and Se have been successfully tested, coupled to a thin Ge wafer operated as a bolometer for the light readout \cite{Pirr06,Arnaboldi:2010jx,Arnaboldi:2010tt,Beeman2012318,Gironi:2010hs,ZnMoO4_1,ZnMoO4_2}. At present scintillating crystals that look most promising for a large scale \BBz experiment are ZnSe, \tungstato, and \zincato. 

%
%

A recent discovery in the field of scintillating bolometers~\cite{Arnaboldi:2010gj} opens up a new analytical technique to increase the background rejection power. When the bolometer is a scintillator, the heat signal presents a different time-dependent shape according to the amount of energy that flows into non-radiative processes of the light channel~\cite{Gironi:PSD}. This allows discrimination of $\alpha$s from the $\beta/\gamma$ particle populations without the light readout. The ability to recognize the interacting particle from the different pulse shapes of the thermal signal was demonstrated both for \zincato\ (see details in references~\cite{ZnMoO4_1,ZnMoO4_2,Arnaboldi:2010gj}) and for ZnSe crystals~\cite{Arnaboldi:2010jx}, while in \tungstato\ only small evidence of this feature was observed.

The sensitivity to pulse shape is currently limited by the readout bandwidth of the Neutron Trasmutation Doped (NTD) sensors. This can likely be improved with low-impe\-dance sensors, e.g. Transition Edge Sensors (TES)~\cite{Angloher:2011uu}.

%
\subsection{\teod bolometers with Cherenkov light readout}\label{sec:Cherenkov}
%
According to the present understanding of \teod crystals, they do not scintillate at bolometric temperatures (\ca 10~mK). However, the many advantages offered by this material in terms of bolometric performances and the high natural isotopic abundance of  $^{130}$Te with respect to other candidate nuclei have provided a strong motivation to pursue another, extremely challenging, option: the readout of the Cherenkov light. According to~\cite{TabarellideFatis:2009zz}, \teod crystals have suitable optical properties to act as Cherenkov radiators, with a threshold for Cherenkov light emission of about 50 keV for electrons and about 400 MeV for $\alpha$ particles. Given the typical energies of $\alpha$ particles emitted in radioactive decays (3-10 MeV), they are below threshold for Cherenkov light emission.  This provides, at least in principle, the possibility of tagging $\beta$/$\gamma$ interactions and rejecting the $\alpha$ ones based on the measurement of Cherenkov light.  Cherenkov light emitted by electrons has only recently been observed in a \teod bolometer~\cite{Beeman:2011yc,Casali:2014Ch} while coupled to the same kind of light detector used for scintillating bolometers. 

The $\alpha$ background rejection capability of this technique is not yet comparable with the one obtained with scintillating bolometers. Nevertheless, the results are encouraging and have inspired further R\&D in an effort to increase the 
$\alpha$ vs.~$\beta$/$\gamma$ separation. This could be done through the optimization of the light collection and the reduction of the noise in the light detector or using a new light detector concept (TES~\cite{Angloher:2011uu} or Luke effect~\cite{Isaila2012160} enhanced bolometers, MKIDs~\cite{CalderJLTP2014}, integrated thermistor, etc.).

%
\subsection{Results on $\alpha$ background discrimination power}\label{sec:DP}
\begin{figure*}
\begin{center}
\begin{overpic}[width=0.48\textwidth,clip=true]{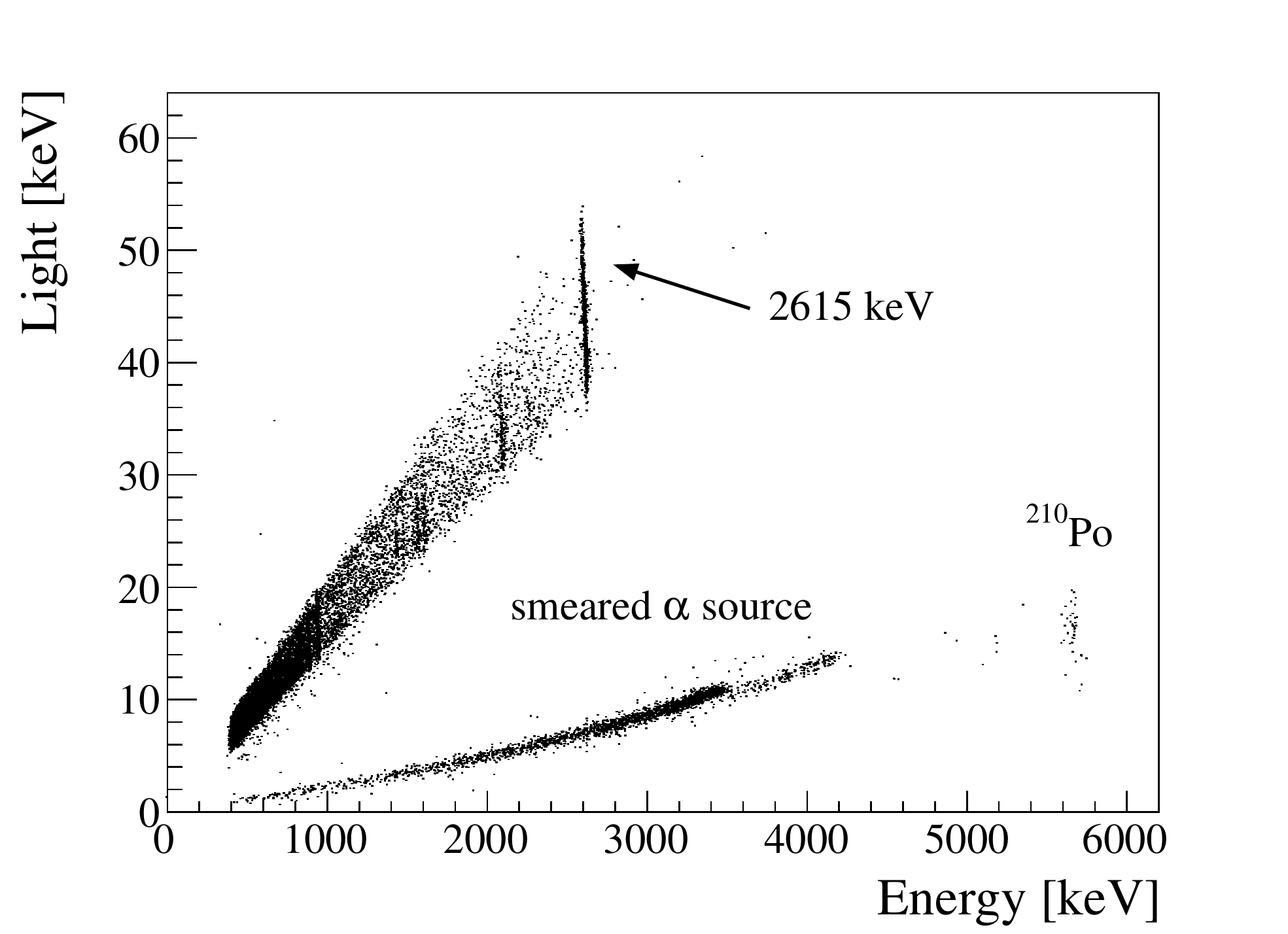}
\put(82,57){\footnotesize a)}
\end{overpic}
\begin{overpic}[width=0.48\textwidth,clip=true]{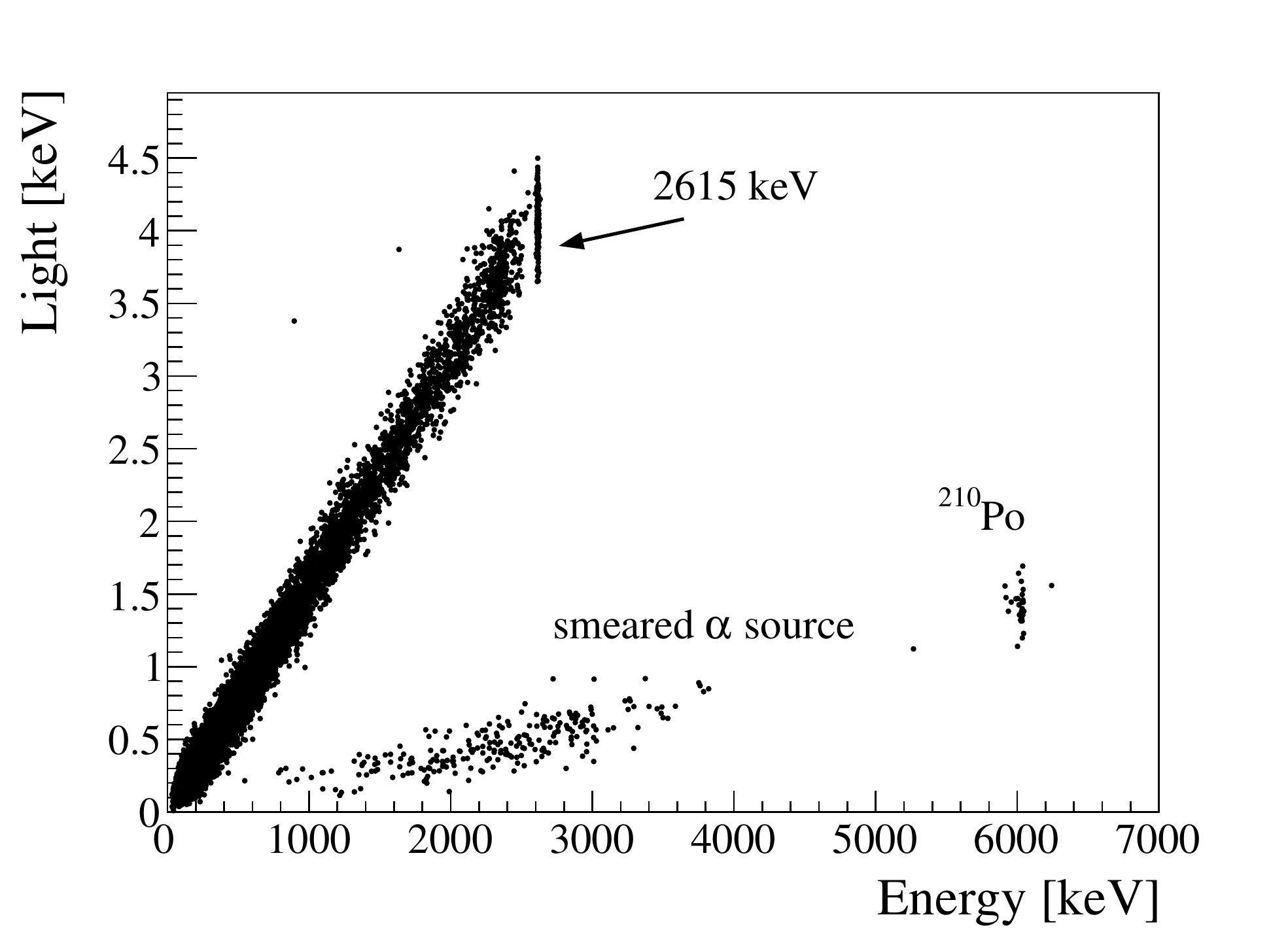}
\put(82,57){\footnotesize b)}
\end{overpic}
\begin{overpic}[width=0.48\textwidth,clip=true]{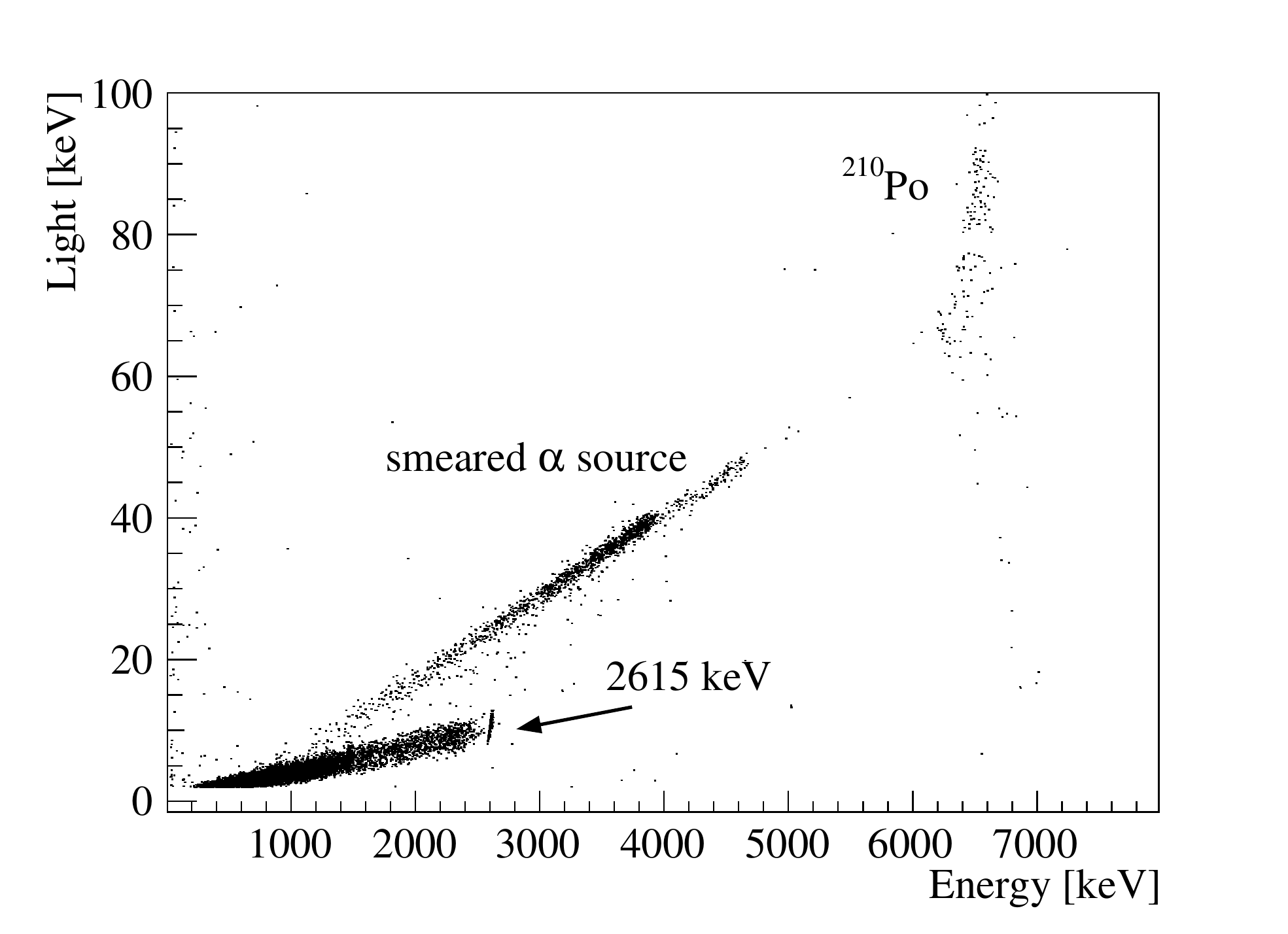}
\put(82,57){\footnotesize c)}
\end{overpic}
\begin{overpic}[width=0.45\textwidth,clip=true]{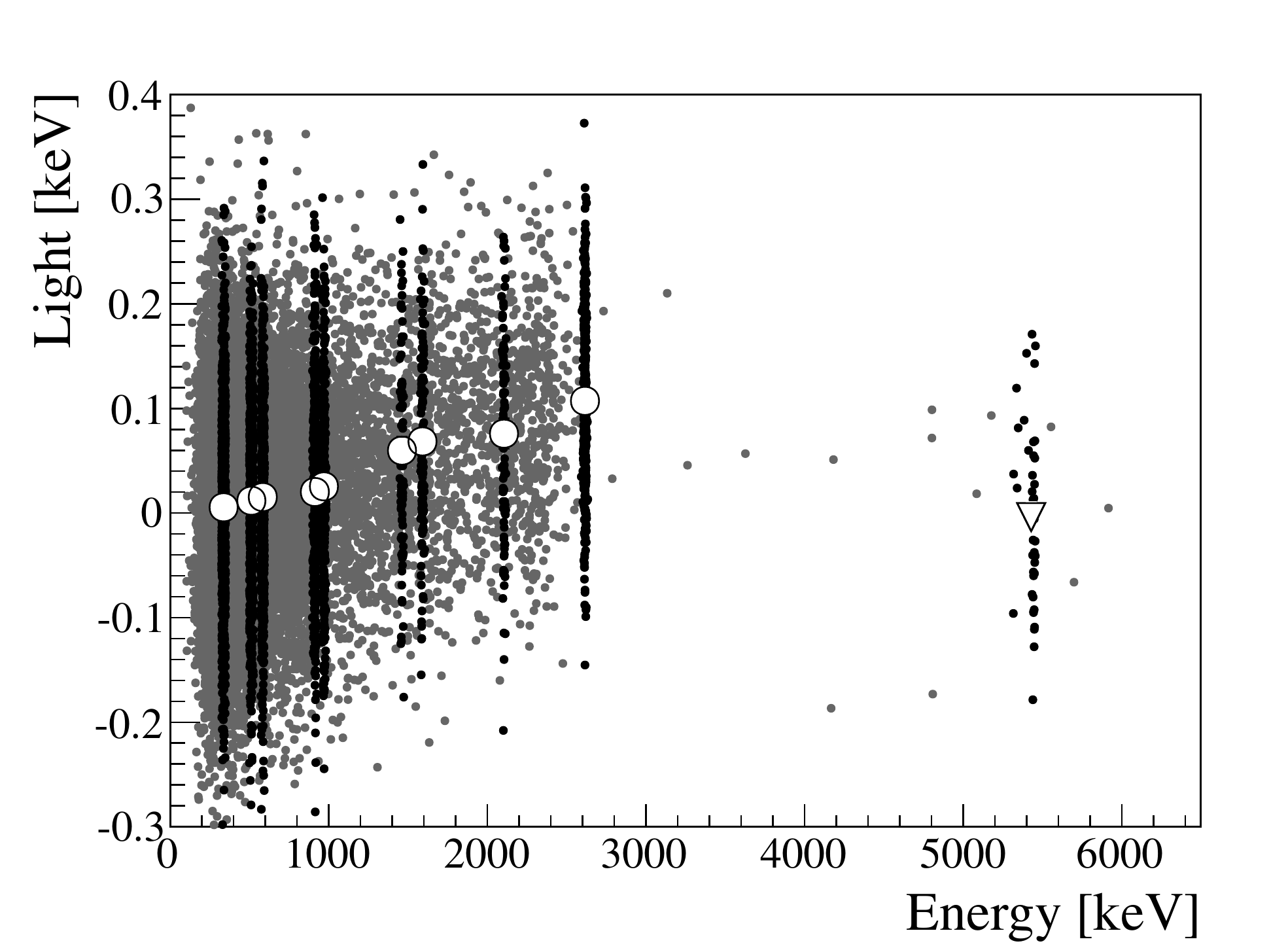}
\put(82,57){\footnotesize d)}
\end{overpic}
\end{center}
\caption{Light vs.~heat scatter plots of recently tested bolometers. The different nature of the light signal should be noted. \tungstato\ (a), \zincato\ (b), and ZnSe (c) are scintillating bolometers where the light signal is given by the scintillation induced by particle interactions. \teod (d) does not scintillate, however Cherenkov light is produced by $\beta$/$\gamma$ interactions (circles) and not by $\alpha$ ones (triangles). Pictures are readapted with authors consensus from the following papers: \cite{Arnaboldi:2010tt} (a), \cite{ZnMoO4_1} (b), \cite{Arnaboldi:2010jx} (c) and \cite{Casali:2014Ch} (d).}
\label{fig:scatter}
\end{figure*}

%
The ability of a scintillating bolometer, or of a bolometer with Cherenkov light readout, to separate $\alpha$ events from $\beta$s and $\gamma$s is demonstrated in Fig.~\ref{fig:scatter} where we have collected the most recent results obtained with ZnSe, \tungstato , \zincato, and \teod .

The bolometers were operated with very similar configurations in a low temperature dilution refrigerator installed underground in Hall C of Laboratori Nazionali del Gran Sasso (LNGS). The $\alpha$ vs. $\beta/\gamma$ rejection factor was determined by exposing the bolometers to a $^{232}$Th calibration $\gamma$ source and to a degraded $\alpha$ source. 
The experimental details regarding each bolometer are described in \cite{Pirr06,Arnaboldi:2010jx,Arnaboldi:2010tt,Beeman2012318,Gironi:2010hs,ZnMoO4_1,ZnMoO4_2,Casali:2014Ch}. As can be seen in Fig.~\ref{fig:scatter} (a,b,c), \tungstato\ and \zincato\ show similar features.  Although they have different light yields, the two classes of particle populations ($\alpha$ and $\beta / \gamma$) are clearly identified. The same is true for ZnSe except the $\alpha$ band lies above the $\gamma$ one. Fig.~\ref{fig:scatter} (d) shows the light versus heat scatter plot recorded in a \teod crystal with Cherenkov light readout. The points belonging to calibration peaks are marked in black and the average light for each peak is shown, both for $\alpha$ peaks (triangles)  and for $\beta / \gamma$ peaks (circles). The Cherenkov light yield for $\alpha$ peaks is compatible with zero, as it should be.

A summary of results concerning the $\alpha$ vs $\beta/\gamma$ discrimination power (DP) is given in Table~\ref{tab:DP}. We define DP as follows:
\begin{equation}
DP = \frac{|\mu_{\beta/\gamma}-\mu_{\alpha}|}{\sqrt{\sigma_{\beta/\gamma}^2+\sigma_{\alpha}^2}} 
\label{eq:DP}
\end{equation} 
where $\mu_i$ is the average value of the distribution of the discriminating parameter for one of the two $i\in\{\beta/\gamma, \alpha\}$ particle populations and 
$\sigma_i$ is the associated width. The discriminating parameter can be the light/heat ratio, a pulse shape variable, or the amplitude of the Cherenkov light signal, according to the given $\alpha$ discriminating technique. 

\begin{table*}
\caption{Summary of the $\alpha$ vs.~$\beta/\gamma$ discrimination power (DP, see Eq.~(\ref{eq:DP})) obtained for several scintillating bolometers and for \teod bolometers.  Results for three techniques are shown: the double readout of the heat and the scintillation light, the pulse shape analysis, and the readout of the Cherenkov light. All results were obtained at the 2615 keV $^{208}$Tl line. For the Cherenkov light readout, the values of the discrimination power reported here are lower than the separations calculated in~\cite{Beeman:2011yc} and~\cite{Casali:2014Ch} due to a different definition of DP.}
\begin{center}
\begin{tabular}{ccccc}
\hline\hline
Bolometer & Scintillation & Pulse shape & Cherenkov & Ref. \\
\hline
\\
ZnSe & 9 & 15 & - & \cite{Arnaboldi:2010jx}\\
\tungstato & 15 & - & - & \cite{Arnaboldi:2010tt}\\
\zincato & 8-17 & 8-20 & -& \cite{ZnMoO4_1,ZnMoO4_2,Arnaboldi:2010gj} \\
\teod & - & - & 1-1.5 & \cite{Casali:2014Ch,Beeman:2011yc}\\
\\
\hline\hline
\end{tabular}
\end{center}
\label{tab:DP}
\end{table*}

\subsection{Energy resolution}
In Table~\ref{tab:risEne} we report the energy resolutions obtained with the scintillating bolometers tested so far as well as the results for the TeO$_2$ crystals. The values in the fourth column (FWHM$_{\theta}$), for scintillating crystals, correspond to an improved energy resolution obtained after correcting for the energy correlation (or anti-correlation) between the heat and light signals (details in~\cite{Beeman:2013znse} and~\cite{Arnaboldi:2010tt}).  

\begin{table*}
\caption{Energy resolution evaluated at 2615 keV for large mass scintillating bolometers tested so far and for \teodn. Resolutions after correcting for the energy correlation (or anti-correlation) between the heat and light signals (details in~\cite{Beeman:2013znse} and~\cite{Arnaboldi:2010tt}) are reported as FWHM$_{\theta}$.}
\begin{center}
\begin{tabular}{ccccccc}
\hline\hline
Crystal ~~~ & mass & ~~~ FWHM ~~~ & ~~~ FWHM$_{\theta}$  & Ref.\\
         & [g] & ~~~ [keV] ~~~ & ~~~ [keV] & \\
\hline
\\
TeO$_2$     & 750 & 5.2 &  & \cite{Alessandria:2011vj}\\ 
ZnSe       & 330 & 28$\pm$1       &   9.5$\pm$0.4 & \cite{Arnaboldi:2010jx} \\
ZnSe       & 431 & 16.3$\pm$1.5       &   13.4$\pm$1.3 & \cite{Beeman:2013znse} \\
\tungstato & 510 & 16.5$\pm$0.5   & 6.25$\pm$0.22 & \cite{Arnaboldi:2010tt}\\
\zincato  &  330 & 6.3$\pm$0.5  &  & \cite{ZnMoO4_2}\\
\\
\hline\hline
\end{tabular}
\end{center}
\label{tab:risEne}
\end{table*}

It is interesting to note the excellent energy resolution obtained by the ZnMoO$_4$ crystals. In fact, for this crystal, the correction for the energy correlation between the heat and light signals was not applied. This feature opens the possibility of an experiment  with discrimination using pulse shape analysis alone, without the need for light detectors to improve the energy resolution.

The resolution quoted for TeO$_2$ crystals was obtained, as explained in~\cite{Alessandria:2011vj}, as the mean value of the resolutions observed in \ciccio CUORE crystals tested in cryogenic runs at the LNGS~\cite{Alessandria:2011vj} in a series of routine tests on the quality of the provided crystals. For the other crystals results reported in Table~\ref{tab:risEne}, one should take into consideration that the results were obtained on test crystals operating under suboptimal noise conditions in a setup not specifically optimized for energy resolution. 
Therefore, in the following discussion, we will make the reasonable assumption that the scintillating crystals, with the optimized thermal design and working conditions, will be able to reach an energy resolution of 5 keV as measured for the \ciccio TeO$_2$ crystals.

%
\section{The Inverted Hierarchy Explorer}
%
The aim of this study is to define the criteria and the constraints for a next generation bolometric experiment able to test the inverted hierarchy region of the neutrino mass spectrum and to assess the sensitivity of such an experiment. In the following we will refer to this experiment as the Inverted Hierarchy Explorer (IHE). 

We will consider as possible choices for the IHE detectors the scintillating bolometers already tested and mentioned above (\zincato, \tungstato, and ZnSe) and \teod bolo\-meters with Cherenkov light readout. 

For each candidate bolometer, we will assume a 90\% isotopic enrichment in the \BB emitting isotope.

\begin{table*}\footnotesize
\caption{IHE characteristics for the different \BB candidates. For each isotope we quote the type of scintillating crystal, the total mass of a 988 \ciccio crystal array, the number of \BB candidates, the number of decays in 5 years (N$_{0\nu\beta\beta}$) for the most and the least favourable values of F$_N$ among those discussed in Section~\ref{sec:NME} for $|m_{ee}|$ = 50 meV and $|m_{ee}|$ = 10 meV. We assume a 90\% isotopic enrichment in the \BB emitting isotope. In the last column we list the 5 year sensitivity at 90\%\,CL under the zero background hypothesis (see Eq.~(\ref{eq:sensitivityzero})).}
\begin{center}
\begin{tabular}{ccccccc}
\hline\hline
Isotope & Crystal & Mass   & N$_{\beta\beta}$ & N$_{0\nu\beta\beta}^{50\rm{meV}}$ &  N$_{0\nu\beta\beta}^{10\rm{meV}}$ & 5 y sensitivity \\
 & & [kg] & & [cnts]  & [cnts] & [y] \\
\hline
\\
 $^{82}$Se & ZnSe   & 664 & 2.4$\times$10$^{27}$      & 10 - 85 & 0.4 - 3.4 & 2.1$\times$10$^{27}$ \\
 $^{116}$Cd & \tungstato& 985 & 1.5$\times$10$^{27}$  & 13 - 44 & 0.5 - 1.8 & 1.5$\times$10$^{27}$ \\
 $^{100}$Mo & \zincato  & 540 & 1.3$\times$10$^{27}$  & 12 - 99 & 0.5 - 4 & 1.1$\times$10$^{27}$ \\
 $^{130}$Te & \teod  & 751 & 2.4$\times$10$^{27}$     & 13 - 89 & 0.5 - 3.6 & 2.5$\times$10$^{27}$ \\
\\
\hline\hline
\end{tabular}
\end{center}
\label{tab:composti}
\end{table*}

\begin{figure*}
\begin{center}
\includegraphics[width=0.5\textwidth,clip=true,angle=-90]{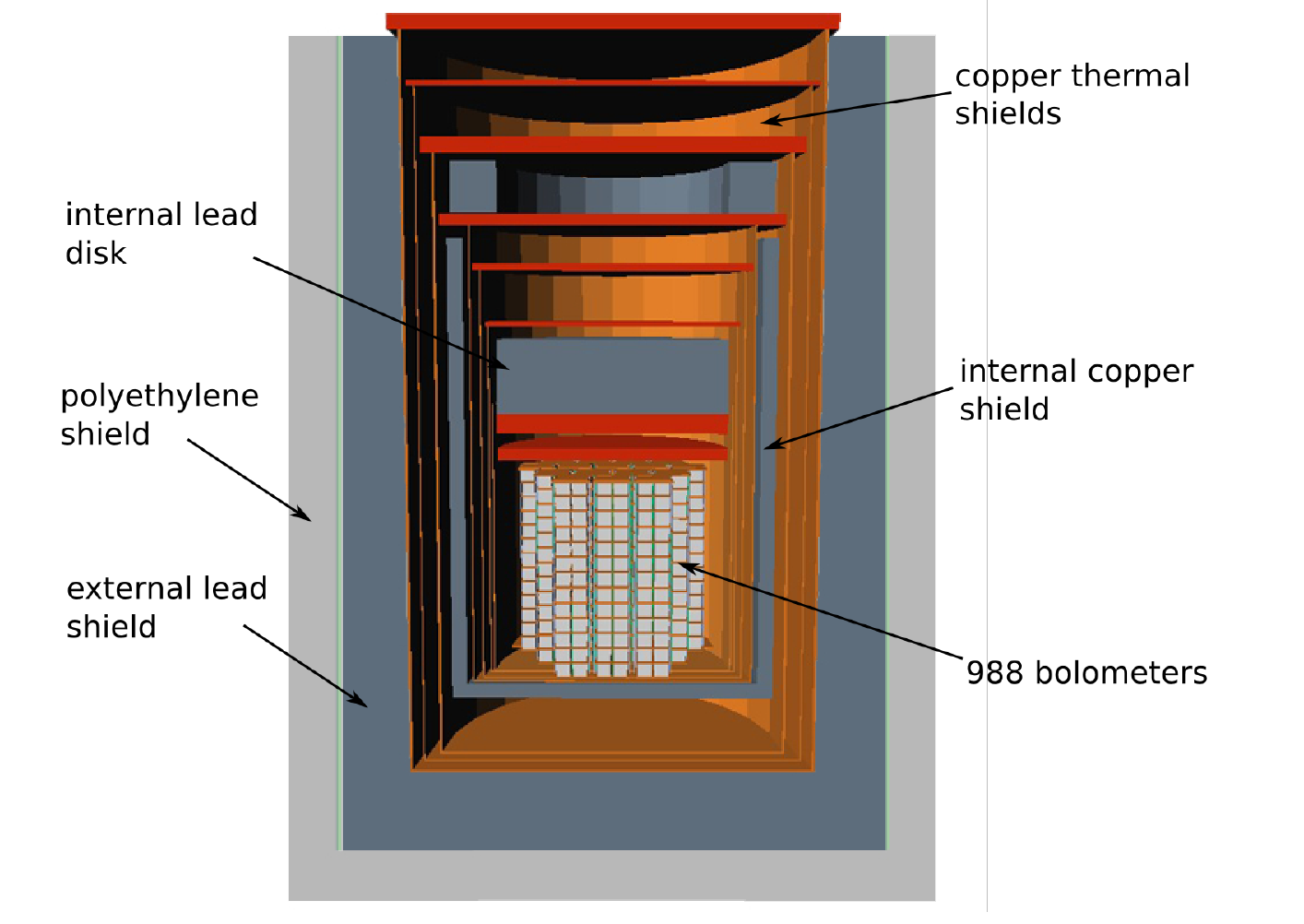}
\end{center}
\caption{Drawing of the IHE geometry, as implemented in the MC simulation: 988 bolometers (\ciccio each) arranged in a close-packed array held by a copper structure, a low temperature refrigerator made of few nested copper thermal shields, a 10~cm thick internal copper vessel shields, a 30~cm thick lead disk placed just above the detector, a 30~cm thick external lead shield, and a 20~cm thick borated polyethylene shield.}
\label{fig:setup}
\end{figure*}

For this discussion, we will assume the IHE experiment will operate underground at LNGS using an experimental set-up similar to the one presently under construction for the CUORE experiment. Fig.~\ref{fig:setup} shows the main elements of the experimental apparatus:
\begin{itemize}
\item \underline{The detector:} 988 bolometers, each with a volume of 125 cm$^3$ (\ciccion), arranged in a close-packed array held by a copper structure. The detectors are arranged in planes of 4 crystals each, with PTFE spacers that secure crystals to the copper frames. Light detectors consisting of Ge or Si ultrapure wafer of about 300 $\mu$m thickness and 10 cm diameter cover each plane. The scintillating crystals are wrapped in a reflecting foil to improve the light collection efficiency.
\item \underline{The cryostat:} a low temperature refrigerator made of few nested cylindrical copper thermal shields. The detector is located inside the inner cylinder. A 10~cm-thick copper vessel shields the bolometers from the radioactivity of the cryostat itself. A 30~cm-thick lead disk, placed just above the detector, provides supplementary shielding against the radioactivity of the various components of the refrigerator located above the detector: the dilution unit, the pumping lines, and the cabling system. These upper components cannot be produced with the same radiopurity required for the thermal shields.
\item \underline{The external shields:} positioned outside the refrigerator, a 30~cm thick lead shield and a 20~cm thick borated polyethylene shield are used to absorb $\gamma$s and neutrons.
\end{itemize}
The main characteristics of the IHE for the different \BB candidates are reported in Table~\ref{tab:composti}.

%
\subsection{Background}\label{sec:bkg}
%
The energy ROI for background evaluation is one FWHM (5 keV) wide and centered at the \BBz Q-value (between 2.528 and 3.035 MeV depending on the isotope). In this region we expect background contributions from the following sources:
\begin{itemize}
\item{Environmental $\mu$s, neutrons, and $\gamma$s;}
\item{\udt and \thdt in setup elements far from the detectors, contributing only through the $\gamma$ emissions of their daughters \bidq and \tldn;}
\item{\udt and \thdt in setup elements close to the detectors, contributing both with their own and their daughters' $\alpha$, $\beta$ and $\gamma$ emissions;}
\item{\BBd and its pile-up in the detectors.  This particular kind of background deserves to be mentioned separately since, for a given enrichment, energy, and time resolution, it is irreducible.}
\end{itemize}

For the intensities of these sources we used either measured values or upper limits, pointing out when further developments are needed in order to reach the desired background level. It should be stressed here that, given the high radiopurity of the selected materials, the sensitivity of the measurement technique plays a crucial role. For example, bulk contamination limits for \udt and \thdt have been obtained with high purity Ge detectors or neutron activation analysis. In the cases where no evidence of contamination was found in the measured samples, the ultimate limit on the background resides in the sensitivity limits of the measuring technique. CUORE itself will provide an high-sensitivity bolometric determination of the radioactive contamination in common detector materials.

Another source of background that should be considered when searching for rare events is the activation of the detector's material and of other materials constituting the experimental setup, by sea-level cosmic-ray neutrons. This process, known as cosmogenic activation, produces long-lived radioisotopes whose subsequent decay can contribute to the background in the ROI. This contribution depends on the amount of isotope produced, that is on the neutron cross section for the given material and on the exposure time at sea level. Among the isotopes produced by cosmogenic activation, only the ones with a Q-value higher that the Q-value of the neutrinoless double beta decay would be contributing to the background in the ROI, so isotopes with higher \BBz Q-value are favored in this respect. The exposure time is also a crucial issue that depends on the history of crystal production. The evaluation of the background induced by cosmogenic activation relies on many complex aspects and it is outside the scope of this paper.

To study the effect of the listed background sources on the background of our experiment, we used a GEANT4-based~\cite{GEANT} Monte Carlo simulation\footnote{The simulation includes the propagation of $\mu$, neutron, $\gamma$, $\beta$, and $\alpha$ particles as well as heavy ions (nuclear recoil after $\alpha$ particle emission). The Livermore physics list is particularly well suited for low energy studies and so it is used here for the generation of nuclear decays or nuclear decay chains.}. The simulation accurately reflects the technical details of the detector geometry, the cryostat, and the internal and external shields. Given the extreme radiopurity and the small thickness, the light detectors and the reflective sheet are not expected to give a sizeable contribution to the background and for this reason they are not included in the simulated geometry. 

In order to study different contributions to the ROI, all the detector elements introduced in the code can serve as a radioactive source with the bulk and surface contamination independently simulated.  In addition, the molecular compound corresponding to the bolometers can be changed in order to study the background expected for the four different detectors. 

The results presented in the following sections, unless otherwise specified, are obtained assuming an $\alpha$ rejection efficiency of 99.9\% and an equal signal selection efficiency.
We do not reject mixed events like the $\beta$+$\alpha$ events produced in the fast decay of the Bi-Po sequence;  due to the slow response of bolometers, events likes these are recorded as single events and their energy deposits are summed.

Moreover, we assume the detectors operate in anticoincidence.  This configuration selects only events where the energy is deposited in a single crystal, as would be the case for a \BBz event. Detailed simulations have shown that the \BBz containment efficiency of a \ciccio crystal depends on the crystal compound and varies from 76\% for ZnSe to 87.4\% for \teod.  Due to the small thickness, the presence of the light detector and reflecting sheet would introduce only minor changes (of the order of few percent) in the anticoincidence efficiency. We neglect these effects at the present level of accuracy. 

%
\subsubsection{Environmental background} 
%
The environmental background at LNGS consists of cosmic ray $\mu$s, neutrons, and $\gamma$s (see for example~\cite{ArtRadio,CUOREExternal} and references therein). The $\gamma$s and neutrons are due to two sources: 1) the natural radioactivity of the rock in the laboratory walls and 2) muon interactions in the rock, materials surrounding the detector, and in the detector itself. The expected background contributions for a CUORE-like experiment, as reported in~\cite{CUOREExternal}, in a ROI of 5 keV around the Q-value are:

\begin{itemize}
\item $\mu$s: \ca $5\times10^{-1}$~cnts/ton/y;
\item neutrons: \ca $5\times10^{-2}$~cnts/ton/y;
\item $\gamma$s: $< 2$~cnts/ton/y (90\% CL). This limit was extracted from an analysis of simulated events with the anticoincidence  selection relaxed due to the low statistics of the simulation.
\end{itemize} 
Further reduction of the environmental background can be obtained by implementing a muon veto and by using thicker lead and polyethylene shields.  In the following we will assume that this contribution can be reduced to negligible levels regardless of the target background counting rate of the experiment. 

%
\subsubsection{Radioactive contamination: Far sources}\label{sec:cryo}
%

We define as far elements all the parts of the experimental setup contributing only $\gamma$s that come from the \thdt and \udt decay chains. Because of the different Q-values, we distinguish between two cases: 1) \teod crystals, for which both \thdt and \udt emissions are relevant, and 2) the other three crystals, for which only \udt contribution is relevant for the background. In fact, for the \thdt decay chain, the only gamma contribution above 2615 keV is given by the summing of this line with other gammas emitted in cascade. The probability of such a summing becomes negligible as the distance of the source from the crystal increases\footnote{In our case, the probability of summing is negligible for the 50 mK thermal shield and farther sources.}.
For the background of the scintillating bolometers we always report the worst result among the three crystals.

The limits on the \udt and \thdt contaminations used for the simulation of far elements have been measured in materials selected for the CUORE experiment. They are reported in Table~\ref{tab:fondoprevisto} together with the resulting background limits from simulations. 

%
\subsubsection{Radioactive contamination: Near sources - bulk}
%

The total amount of material in close vicinity to the crystals is dominated by the copper in the mechanical structure. The resulting bulk background is reported in Table~\ref{tab:fondoprevisto}. 
\begin{table*}\footnotesize
\caption{Background rate induced in a ROI of 5 keV at \BBz by radioactive contamination in several setup elements for an IHE experiment based on \teod crystals or on scintillating bolometers. In the latter case we report the worst result among those obtained for \tungstato, ZnSe, and \zincato. Far Sources correspond to cryostat elements and shields while Near Sources correspond to the detector mechanical structure. The contamination limits for stainless steel were measured in commercially available samples~\cite{Fe-Gerda}. The contamination limits for copper and lead were measured by the authors in specially selected copper (well suited for low temperature applications) and in specially selected lead. Limits are reported at 90\% CL. } 
\begin{center}
\begin{tabular}{lcccc}
\hline
\hline
Element	& material & contamination  & Te & Se/Cd/Mo \\
& & [Bq/kg] & \multicolumn{2}{c}{[cnts/ton/y]} \\
\hline
\multicolumn{5}{c}{Far Sources} \\
\hline
\\
\udt  external shield & lead 			& $< 1\times10^{-5}$ 	&  $< 7\times10^{-3}$ 	&  $< 4\times10^{-3}$\\
\thdt  external shield  & lead 			& $< 7\times10^{-5}$ 	&  $< 1$	 	&  $< 1\times10^{-2}$\\
\udt  300 K top plate & stainless steel 	& $< 2\times10^{-4}$ 	&  $< 5\times10^{-4}$	&  $< 3\times10^{-4}$\\
\thdt 300 K top plate & stainless steel 	& $< 1\times10^{-4}$ 	&  $< 3\times10^{-2}$ 	&  $< 3\times10^{-4}$\\
\udt cryostat elements & copper 		& $< 7\times10^{-5}$ 	&  $< 4\times10^{-1}$	&  $< 3\times10^{-1}$\\
\thdt cryostat elements & copper 		& $< 2\times10^{-6}$ 	&  $< 3\times10^{-1}$ 	&  $< 1\times10^{-2}$\\
\udt internal shield & copper 			& $< 7\times10^{-5}$ 	&  $< 1$		&  $< 6\times10^{-1}$\\
\thdt internal shield & copper 			& $< 2\times10^{-6}$ 	&  $< 8\times10^{-1}$	&  $< 8\times10^{-3}$\\
\udt 30~cm disk & lead 				& $< 1\times10^{-5}$ 	&  $< 1\times10^{-3}$	&  $< 7\times10^{-4}$\\
\thdt 30~cm disk & lead 			& $< 7\times10^{-5}$	&  $< 2\times10^{-1}$	&  $< 2\times10^{-3}$\\
\\
\hline
\multicolumn{5}{c}{Near Sources} \\
\hline
\\
\udt detector holders & copper 			& $< 7\times10^{-5}$ 	&  $< 2$	&  $< 1$\\
\thdt detector holders & copper 		& $< 2\times10^{-6}$ 	&  $< 1\times10^{-1}$	&  $< 2\times10^{-1}$\\
\\
\hline
\hline
\end{tabular}
\end{center}
\label{tab:fondoprevisto}
\end{table*}

Table~\ref{tab:bulk} summarizes the crystal bulk contamination results obtained in dedicated underground tests where the crystals were operated as bolometers. The contamination limits for \teod have been measured in a random sample of crystals produced for the CUORE experiment, which followed very strict radiopurity protocols. The upper limits for the other, scintillating, bolometers were obtained with prototypes grown without rigorous attention to material selection. 

The dominant contribution to the background per unit source intensity simulated is from $\beta$/$\gamma$ emissions of \tldn, \bidq, and \pbddn. Further background reduction with respect to the values reported in Table~\ref{tab:bulk} can be obtained by exploiting the technique of delayed coincidences. For example, \tld $\beta$-decays with a Q-value of 5 MeV and a half-life of 3 minutes. The background induced by this isotope can be rejected with the use of a delayed coincidence between the \tld signal and the $\alpha$ emitted by its precursor, \bidd (E$_{\alpha}$ = 6 MeV). 
The choice of the coincidence window width is a compromise between the background reduction factor (F$_{\rm{B}}$) and the resulting dead-time (T$_{\rm{dead}}$). For  \thdt contamination of 8$\times 10^{-6}$~Bq/kg (the worst case reported in Table~\ref{tab:bulk}), the compromise sets F$_{\rm{B}}$\ca3 and T$_{\rm{dead}}$\ca10\%. With one order of magnitude lower contamination, which may be possible with a dedicated material purification campaign, we obtain F$_{\rm{B}}$\ca20 and T$_{\rm{dead}}$\ca3\%.

\begin{table*}
\caption{Crystal bulk contamination levels and the corresponding background counting rate in the ROI (without rejection by delayed coincidence, see text). Here the background is dominated by $\beta$/$\gamma$ events.  Limits are reported at 90\% CL.  } 
\begin{center}
\begin{tabular}{lccccc}
\hline
\hline
\\
Crystal & \udt     & \thdt    &  \udt in ROI  & \thdt in ROI  & Ref.\\
        &  [Bq/kg] &  [Bq/kg] &  [cnts/ton/y] &  [cnts/ton/y]  &  \\
\hline        
\\
\teod   	& $< 7\times10^{-7}$ 		& $< 8\times10^{-7}$ & $< 2\times10^{-2}$ 	& $< 5\times10^{-1}$ & \cite{Alessandria:2011vj} \\
ZnSe    	& $< 4\times10^{-7}$ 		& $< 4\times10^{-7}$ & $< 3\times10^{-2}$ 	& $< 3\times10^{-1}$ & \cite{Arnaboldi:2010jx} \\
\tungstato\ 	& $< 4\times10^{-5}$ 		& $< 4\times10^{-6}$ & $< 1$	 		& $< 5$ & \cite{Arnaboldi:2010tt} \\
\zincato\ 	& $(27\pm6)\times10^{-6}$ 	& $< 8\times10^{-6}$ & $(5.5\pm1.0)\times10^{-1}$ & $< 5$ & \cite{ZnMoO4_2}\\
\\
\hline
\hline
\end{tabular}
\end{center}
\label{tab:bulk}
\end{table*}

Also, \bidq $\beta$-decays with a branching ratio of 99.98\% to $^{214}$Po, which in turn $\alpha$-decays with a very short half-life (163 $\mu$s) and a Q-value of 7.8 MeV. Bolometer signals develop over about 2 seconds, therefore the chain \bidq $\to^{214}$Po$\to^{210}$Pb gives rise to a pile-up event that is easily rejected. In this case, the energy released by the two decays adds together and generates a continuous background whose lower limit is the Q-value of the $^{214}$Po $\alpha$ decay. 
This is already taken into account in the simulation.
In the remaining 0.02\% of the cases, \bidq $\alpha$-decays to \tldd, which is a $\beta$ emitter with a Q-value of 5.4 MeV and a half-life of 1.3 minutes. The background induced by this $\beta$ emitter can be rejected  with the use of a delayed coincidence between the \tldd signal and the \bidq signal.

As an additional remark about this kind of background we discuss the contribution of $^{113}$Cd in \tungstato\ crystals.
This $\beta$-decaying isotope (Q-value=316~keV) is responsible for the high natural activity of \tungstato\ crystals (\ca0.5 Bq/kg in natural \tungstato) and could contribute to the ROI counting rate through spurious pile-up events. 
However, isotopic enrichment in the \BB candidate $^{116}$Cd will deplete $^{113}$Cd reducing the pile-up-induced background.  Enrichment in this case will also decrease the rate of (n,$\gamma$) reactions on $^{113}$Cd, whose neutron cross section is extremely high.

Finally, bulk contaminations in the light detectors can be neglected because Ge and Si wafers, whose masses are only on the order of a few grams, are generally characterized by a high intrinsic radiopurity.  Also, events occurring within the light detectors can be easily tagged and rejected. For example, given the typical energies of $\alpha$ and $\beta$ decays, the energy deposition in the light detector will be much higher than the ones typically produced by scintillation light.  Moreover, a direct particle event within the light detector can also be easily rejected by pulse shape analysis.

%
\subsubsection{Radioactive contamination in near elements: surface}
\begin{figure}
\begin{center}
\includegraphics[width=0.48\textwidth,clip=true]{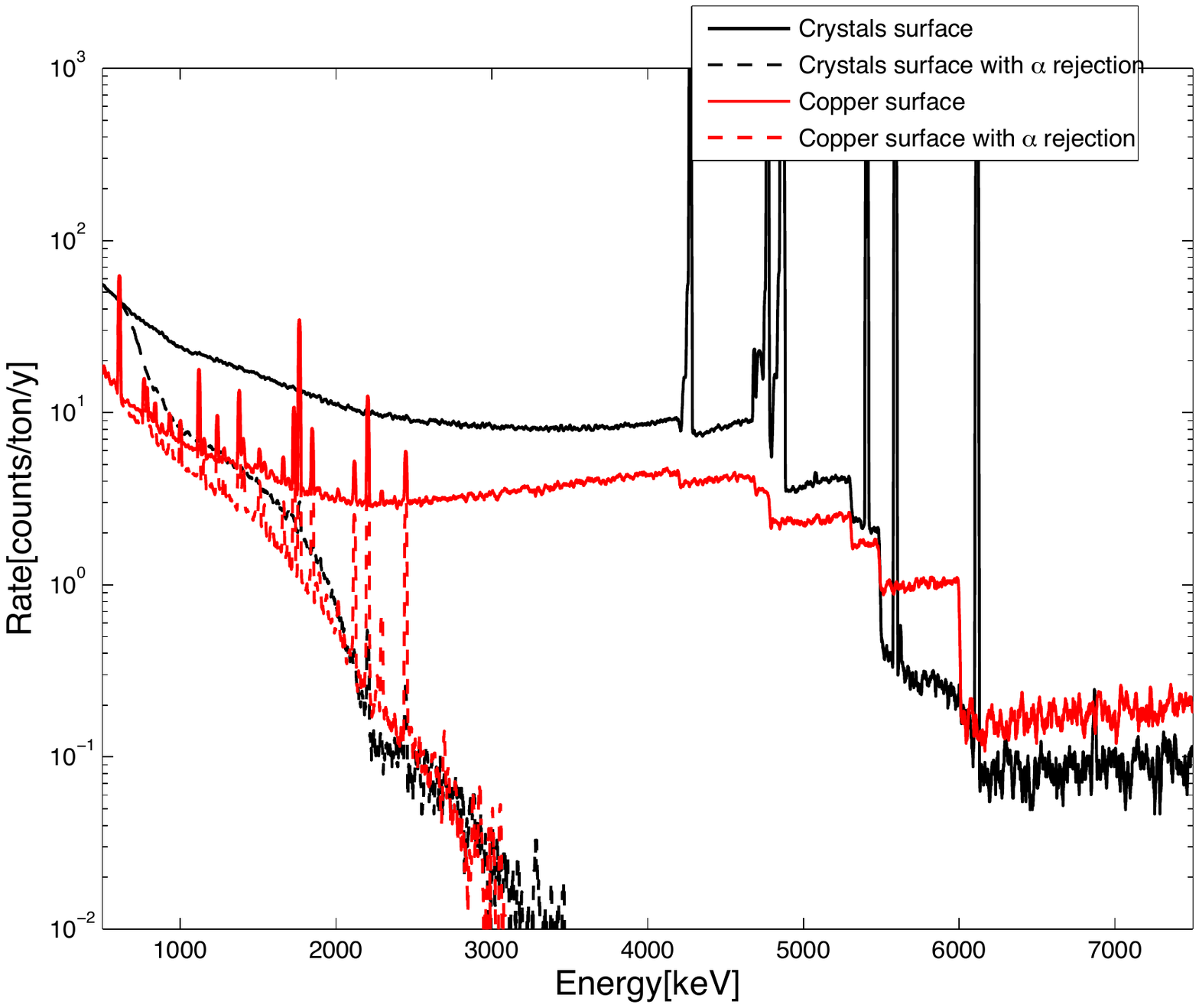}
\includegraphics[width=0.48\textwidth,clip=true]{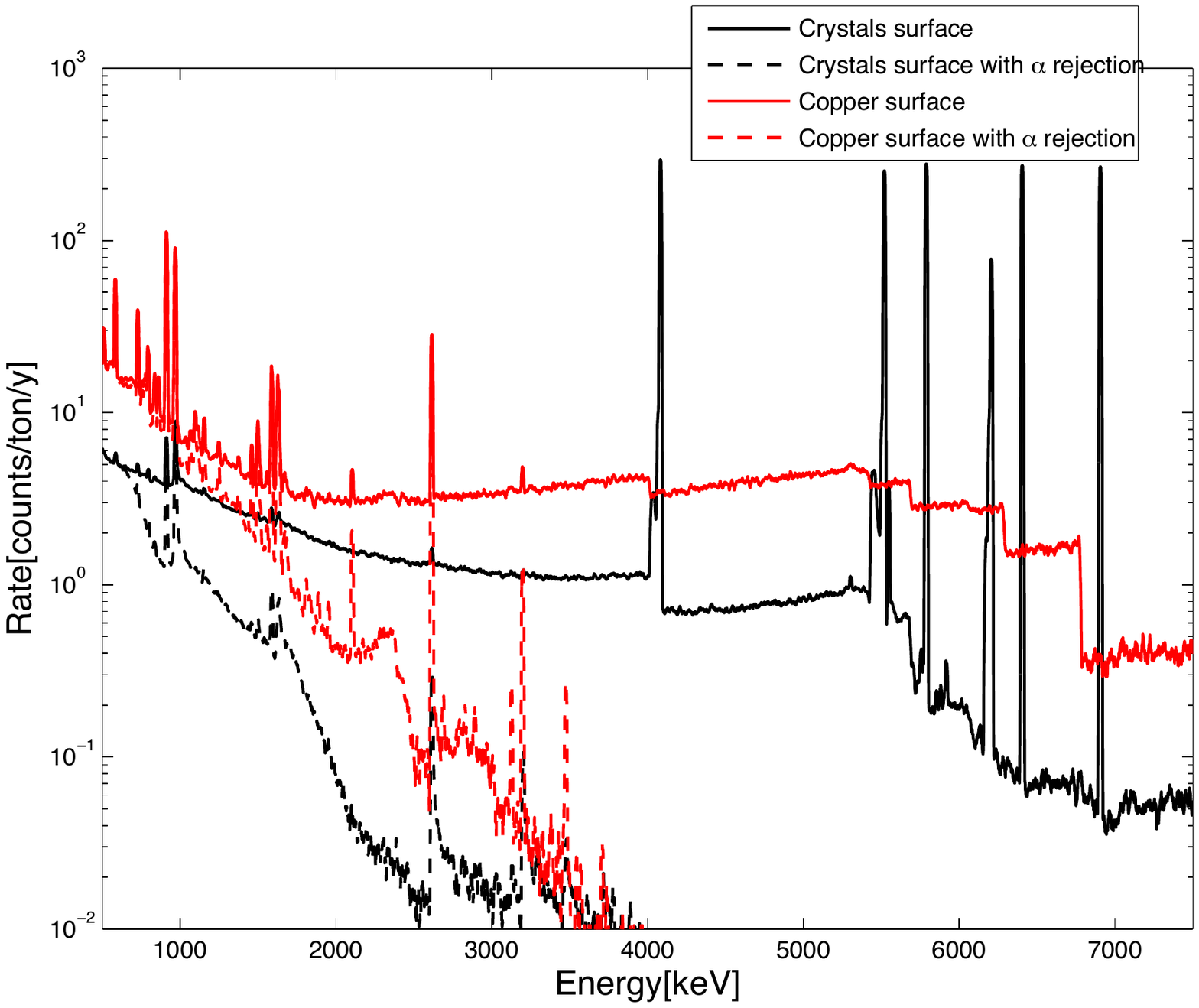}
\end{center}
\caption{Effect of $\alpha$ background rejection for \udt (top) and \thdt (bottom) contaminations on \teod crystal surfaces and copper surfaces. The spectra have been obtained simulating a surface contamination with an exponential depth profile and the mean depth of 5 $\mu$m. Solid histograms correspond to the total simulated background contribution from the given source (crystal or copper) while in the dashed histograms the $\alpha$ contribution is removed.}
\label{fig:alpha_rejection}
\end{figure}

%
It is often observed, particularly in the most radiopure materials, that surface contamination exceeds the contamination coming from the bulk.  This is generally due to the external mechanical and/or chemical treatment of the materials or because the materials are exposed to contaminated air. When considering elements that are close to the bolometers, these contaminants can be problematic in limiting sensitivity. For example, in the Cuoricino experiment a large fraction of the counting rate in the ROI was attributed to surface contamination on the detector materials~\cite{Andreotti:2010vj};  this surface background source will most probably be the one limiting CUORE sensitivity~\cite{ACryo,CuoreBB}. The $\alpha$ rejection capability of scintillating bolometers is particularly effective in reducing this background source. Fig.~\ref{fig:alpha_rejection} illustrates the effect of $\alpha$ background rejection for \udt (left) and \thdt (right) contaminations on \teod crystal surfaces and copper surfaces. 

The impact of surface contamination on the background depends critically on the contaminant (\udt, \thdt, $^{210}$Pb) and on the depth of the contaminated layer. Both are generally difficult to measure~\cite{Alessandria:2011vj,ArtRadio,ArtRadio-CUORE,koglerphd}.  Therefore, we assume a reasonable set of values based on our past experience with bolometers. The results reported in Table~\ref{tab:surface} are obtained assuming, for both the crystals and the copper in contact with them, a surface contamination with an exponential depth profile and the mean depth of 5 $\mu$m\footnote{The real depth of the \teod crystal surface contamination is not exactly known.  However, we have used the most conservative value among those analyzed in~\cite{Alessandria:2011vj}, that is the one which produces the highest background in the ROI.}. The measured activities (or limits) of the various contaminants are also reported. For crystal surfaces, the activity limits were measured in \teod bolometers~\cite{Alessandria:2011vj} and no evidence for surface contamination was found. The extreme radiopurity of the surfaces has been achieved with a dedicated protocol of polishing and cleaning that can be reasonably applied (and most probably improved) to \tungstato\, ZnSe, and \zincato\ as well. 

For the copper surface contamination, we refer to the results obtained in~\cite{CUORE0Paper} for the CUORE-0 experiment, a single CUORE-like tower and a technical prototype of CUORE, operated at LNGS since March 2013. Only indirect evidence of surface contamination was found\footnote{This means that a given event rate was observed in several energy regions compatible with the existence of surface contamination, but no distinct signature could be attributed directly to a particular source of contamination.}. We use the same procedure as in~\cite{TTT:2013ap} to convert the measured rate into a quantitative limit on the surface contamination, that is we assume that the entire rate measured in~\cite{CUORE0Paper} is due to each species of impurity on the copper surface (\udt or \thdt or $^{210}$Pb) in turn. Therefore, the derived surface contamination values are conservative and set mutually exclusive upper limits.

We recall that the background rates shown in Table~\ref{tab:surface} are obtained assuming a 99.9\% rejection of $\alpha$-induced counts and an anticoincidence cut among the detectors. Surface contaminations in the light detectors are not considered because they would give rise to coincident events, which are easily tagged and rejected.

\begin{table*}
\caption{Upper limits (90\% CL) on surface contamination of crystals and copper (see text for more details) with the corresponding upper limit on the induced \BBz counting rate in a ROI of 5 keV. A 99.9\% rejection of $\alpha$-induced counts and anticoincidence cuts among the detectors have been applied.} 
\begin{center}
\begin{tabular}{lccc}
\hline
\hline
Element	& Contamination & Te & Se/Cd/Mo \\
& [Bq/cm$^2$] & [cnts/ton/y] & [cnts/ton/y] \\
\hline
\\
\udt on crystal surface 	& $< 9\times10^{-9}$ 	&  $<2$ & $< 1$ \\
\thdt on crystal surface  	& $< 2\times10^{-9}$ 	&  $<1\times10^{-1}$ 	& $< 2\times10^{-1}$  \\
\pbdd on crystal surface  	& $< 2\times10^{-8}$ 	&  $<8\times10^{-3}$ 	& $< 9\times10^{-3}$ \\
\\
\hline
\\
\udt on copper surface 		& $< 3\times10^{-8}$ 	&  $<1$		&  $< 3\times10^{-1}$\\
\thdt on copper surface	 	& $< 4\times10^{-8}$ 	&  $<1$			&  $< 2$\\
\pbdd on copper surface  	& $< 2\times10^{-7}$ 	&  $<3\times10^{-2}$ 	&  $< 4\times10^{-2}$ \\
\\
\hline
\hline
\end{tabular}
\end{center}
\label{tab:surface}
\end{table*}

%
\subsubsection{\BBd induced background}
%
The background sources described above (external sources or radioactive contaminants in the setup) could theoretically be reduced to zero, even though the technical challenges would be substantial.  However there is a background source for any \BBz search that is always present: the \BBd decay of the candidate isotope itself.

The end point of the \BBd spectrum can contribute substantially to the background in the ROI as the energy resolution of the experiment becomes larger. However, as already shown in Table~\ref{tab:isotopi}, in detectors with good energy resolutions like bolometers, the ratio of \BBd to \BBz event rate, assuming $|m_{ee}|$ in the IH region, is negligible.

On the other hand, a drawback in the use of bolometers comes from their slow response time.  Accidental pile-up of \BBd events can produce a contribution to background in the ROI at a detectible level, thus limiting the sensitivity of an experiment~\cite{Beeman2012318,Chernyak:2012}. Two events produced in the same detector by two random \BBd decays within a time window smaller than the typical time response of the detector can produce a signal that mimics a \BBz decay. 

In order to study the effects of \BBd pile-up in bolometric detectors in detail, we used a software tool developed for the CUORE experiment~\cite{Carrettoni:2011rn} that simulates signal pulses and noise samples of \teod bolometers, including the effects generated by operating temperature drifts, nonlinearities, and pile-up. 
The signal shape is reproduced by means of a thermal model~\cite{Vignati:2010yf} with a pulse amplitude randomly extracted from a theoretical \BBd spectrum \cite{zuber-libro}. The pulse is then superimposed on a noise baseline, sampled according to measured noise power spectra of real \teod detectors.
The pile-up rate is artificially increased so that two \BBd pulses always pile-up within a time window between 0 and 100 ms. The simulated pile-up pulses are then processed as real data. 

The analysis shows that standard pulse shape analysis cuts, already developed for CUORE,  give a pile-up rejection efficiency of 100\% down to $\Delta$T = 5 ms and, for the best performing channel, down to  $\Delta$T = 1 ms. In Table~\ref{tab:pile-up} we summarize the \BBd pile-up-induced background for the four \BB candidates considered in this paper. The results for scintillating crystals are an extrapolation of what was obtained for \teod crystals using the simulation framework described above. The pile-up rate is evaluated for 90\% enrichment, a \ciccio crystal, and for a minimum pulse separation times of $\Delta$T = 1 ms. The background is expressed in cnts/ton/y in a ROI of 5 keV.

Further improvement could be obtained by exploiting the faster time response of light detectors;  the pulse rise time of bolometric light detectors is already a factor $\sim$5 smaller than the rise time of bolometric signals. For example, the case of $^{100}$Mo, which is the most problematic among the isotopes in Table~\ref{tab:pile-up}, given the relatively small T$_{1/2}^{2\nu}$, has been extensively studied in~\cite{Chernyak:2012}.  There, it is shown that the pile-up discrimination on light signals can reduce the background induced by \BBd in $^{100}$Mo to well below 1 cnts/ton/y.
 
\begin{table*}
\caption{\BBd pile-up induced background in a 5 keV ROI for experiments based on the four \BB candidates discussed in this work. The pile-up rate is evaluated for a 90\% enrichment, a \ciccio crystal and for a minimum pulse separation time $\Delta$T = 1 ms.}
\begin{center}
\begin{tabular}{ccccc}
\hline\hline
Isotope & Crystal & N$_{\beta\beta}$ & T$_{1/2}^{2\nu}$ & Bkg in ROI [5 keV] \\
 & & [n/crystal] & [y] & [cnts/ton/y] \\
\hline
\\
 $^{82}$Se  	& ZnSe  		& 2.5$\times$10$^{24}$  	& 9.2$\times$10$^{19}$ & 2.7$\times$10$^{-2}$ \\
 $^{116}$Cd 	& \tungstato	& 1.5$\times$10$^{24}$  	& 2.8$\times$10$^{19}$ & 0.07 \\
 $^{100}$Mo 	& \zincato  	& 1.3$\times$10$^{24}$  	& 0.7$\times$10$^{19}$ & 1.5  \\
 $^{130}$Te 	& \teod  		& 2.5$\times$10$^{24}$  	& 68$\times$10$^{19}$  & 0.5$\times$10$^{-3}$ \\
\\ 
\hline\hline
\end{tabular}
\end{center}
\label{tab:pile-up}
\end{table*}

In principle a pile-up can occur not only among two \BBd decays but also among two radioactive background pulses or among a \BBd decay and a radioactive decay. Both these additional contributions can result in spurious pulses in the ROI. While the \BBd pile-up rate, given the isotope and the crystal, is irreducible, the pile-up rate from background and background plus \BBd depends on the radioactive background rate and can therefore be limited to a certain extent.
Our estimations show that, assuming a radioactive background rate of 0.14 mHz (CUORE estimate), the resulting pile-up background in the ROI from radioactive sources and radioactive sources plus \BBd is negligible for all isotopes but for Tellurium, where the contribution is of the same order of magnitude than the \BBd pile-up. However, considering the additional background reduction foreseen for the IHE with respect to CUORE, we conclude that the only source of pile-up background in the ROI that has to be taken into account for a future generation bolometric experiment like the IHE is the \BBd decay, as considered in this work.
%
\subsection{Background budget}
\label{sec:budget}
\begin{figure}
\begin{center}
\includegraphics[width=0.5\textwidth,clip=true]{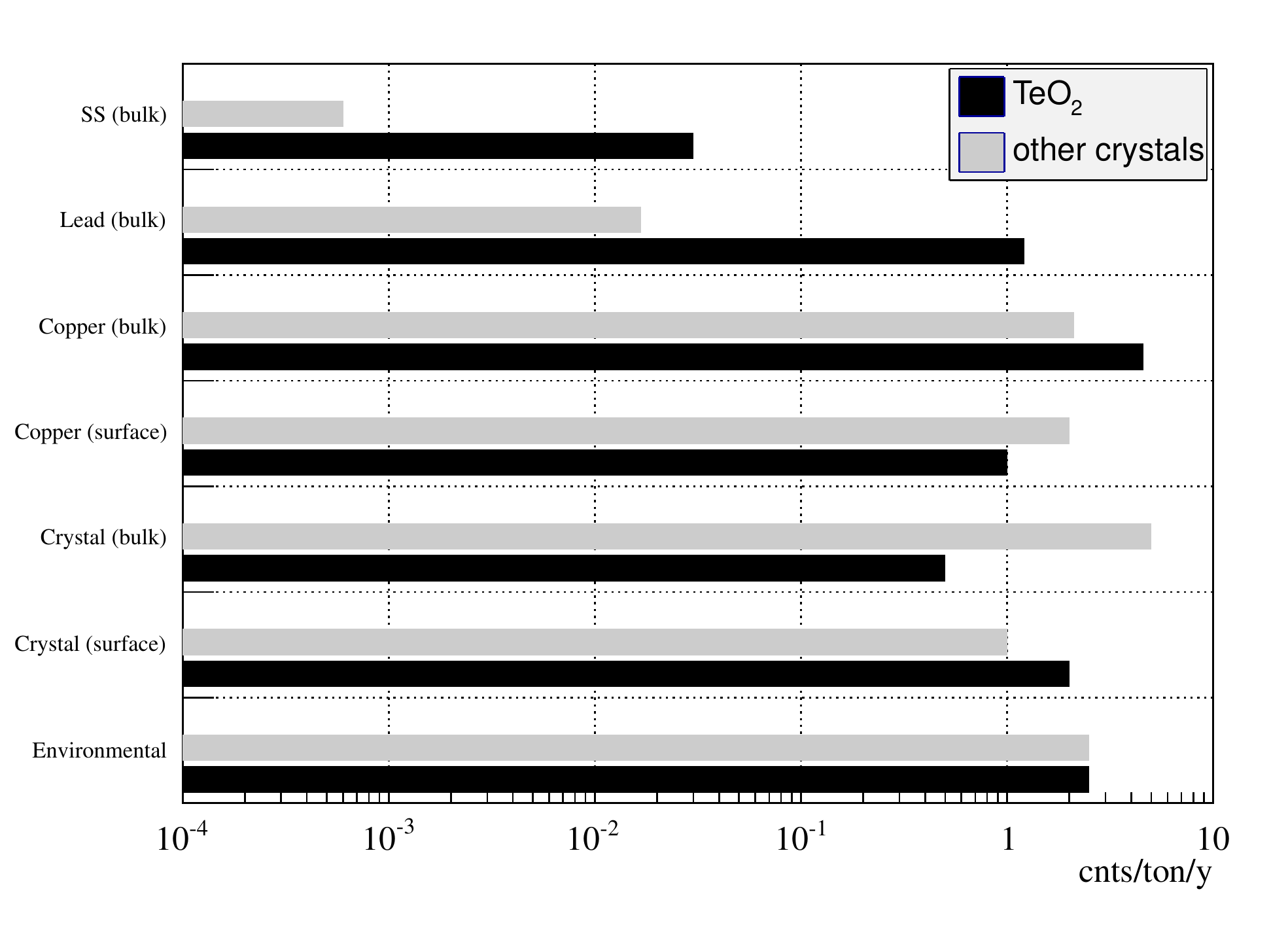}
\end{center}
\caption{IHE background budget. All reported values are conservative upper limits. Black bars refer to \teod crystals while grey bars refer to other crystals. \BBd pile-up-induced background is not included. SS stands for stainless steel.}
\label{fig:budget}
\end{figure}

%
Here, we summarize the results that are described in previous sections and reported in Tables~\ref{tab:fondoprevisto}, \ref{tab:bulk}, and \ref{tab:surface};  the tentative background budget of the IHE is shown in Fig.~\ref{fig:budget}. The different background contributions are grouped according to the material where the contamination is located. Upper limits for \udt and \thdt for the same material were summed up in order to give the most conservative result, except for copper surface contamination (see Table~\ref{tab:surface}) where the upper limits for different isotopes were not summed since they are mutually exclusive. In this case, the highest upper limit was taken. For the materials where both the bulk and surface contaminations contribute to the background in the ROI, the two components are indicated. Black bars refer to \teod crystals while grey bars refer to scintillating crystals.  For the latter, we indicate the worst result among those available. \BBd pile-up-induced background is not included in the budget of Fig.~\ref{fig:budget}. 

As indicated previously, all reported background values are conservative upper limits which reflect the state-of-the-art of the ongoing R\&D including: scintillating bolometers, material cleaning techniques, and methods for measuring such low levels of radioactive contamination.  Low level radioactivity measurements in particular are becoming more and more challenging as the radiopurity of the materials increases. It is reasonable to expect that all the reported limits will improve in the near future. CUORE-0 initial performances~\cite{CUORE0Paper} have already demonstrated a factor of 6 improvement in the background rate in the $\alpha$ continuum region compared to Cuoricino. This is due to more rigorous copper surface treatment, improved crystal production and treatment protocols, as well as more stringent assembly procedures. CUORE-0 further data and CUORE itself will provide valuable measurements of the radiopurity ultimately achieved with state of the art materials cleaning techniques.

In order to draw a conclusion on what could be the ultimate background reach of a IHE bolometric experiment, we make the aggressive assumption that using a muon active veto and a neutron shield, applying to the scintillating crystals the same (or even more stringent) protocols of material selection and crystals production and polishing used for \teod crystals, using delayed coincidences to tag $\beta/\gamma$ bulk emissions, and understanding the nature of surface contaminations, a background index of 0.1 cnts/ton/y could be optimistically achieved.

%
\subsection{Discovery potential}
%
In the following discussion, we will refer to the definition of sensitivity and discovery potential given in Ref.~\cite{CUOREsensi} for the CUORE experiment. 

In order to claim \BBz discovery, the exposure of the experiment (e.g. measured in ton$\cdot$y) should be large enough to generate a statistically significant number of signal events above the background. When a signal is observed in the presence of a finite background, there is always some probability that the observation is due to a background fluctuation. The convention for discovery is that any candidate signal be greater than 5$\sigma$ above the background, corresponding to a probability of $\alpha = 2.87 \times 10^{-7}$ that the observation is merely a fluctuation consistent with the expected background. In this case the finite-background Gaussian-regime discovery potential is defined by Eq.~(\ref{eq:sensitivity}) with $n_{\sigma} =5$. 

For extremely low background levels, such as those considered for the IHE in this work, the Gaussian regime no longer holds and it is more appropriate to assume a Poisson distribution of the background counts. Following~\cite{CUOREsensi}, the Poisson-regime calculation of the background-fluctuation sensitivity is given by the formula:
\begin{equation}
P(\widehat{S}(\Delta E) + B(\Delta E),B(\Delta E)) = \alpha~,
\label{eq:PoissonSens}
\end{equation}
where $\widehat{S}(\Delta E)$ and $B(\Delta E)$ are the expected number of signal and background events in an energy region $\Delta E$ around the Q-value, which is in our case 5 keV. The quantity $\widehat{S}(\Delta E)$ is defined as:

\begin{equation}
\widehat{S}(\Delta E) =  \widehat{S}_0 f(\Delta E) ~,
\label{eq:PoissonSig}
\end{equation}
where $\widehat{S}_0$ is the mean signal and $f(\Delta E)$ is the fraction of signal events that fall in an energy window $\Delta E$ around the Q-value. Eq.~\ref{eq:PoissonSens} means that the Poisson integrated probability that the background distribution alone will cause a given experiment to observe a total number of counts larger than $\widehat{S}(\Delta E) + B(\Delta E)$ is lower that $ \alpha $.

The Poisson finite-background discovery potential can be defined as the \BBz half-life that would give rise to the mean signal $\widehat{S}_0$ found from Eq.~(\ref{eq:PoissonSens}) for  $\alpha = 2.87 \times 10^{-7}$ and the appropriate value of  $B(\Delta E)$.

In the true zero-background case, the observation of a single event would be enough to claim discovery.  However, it is still necessary to quantitatively set the criteria to see that one event. A conservative choice is to require $\widehat{S}(\Delta E)$ be high enough so that the Poisson-regime probability to observe at least one event is higher than 90\%. This corresponds to $\widehat{S}(\Delta E) = 2.3$. Therefore we can set the two criteria for discovery potential as: 
\begin{enumerate}
\item $ P(\widehat{S}(\Delta E) + B(\Delta E),B(\Delta E))\leq 2.87 \times 10^{-7} $
\item $\widehat{S}(\Delta E) \geq 2.3 $
\end{enumerate}
For a sufficiently small number of background counts, the requirement for the expected observation to be inconsistent with the background (criterion 1) becomes less stringent than the requirement for the experiment to be reasonably likely to observe any signal event at all (criterion 2). 

To discuss the discovery potential of the IHE, we defined it as a CUORE-like (988 \ciccio crystals) double beta experiment using bolometer technology based on the four \BBz candidates discussed in this work. Further assumptions are: 90\% enrichment in the candidate isotope, $\Delta E$= 5 keV, and a 5 y exposure time. For the background, we set it to either 0.1 cnts/ton/y (discussed in Sec.~\ref{sec:budget}) or to the value of the 2$\nu\beta\beta$ pile-up-induced background for $\Delta T $= 1 ms (see Table~\ref{tab:pile-up}), whichever is bigger.

In Table~\ref{tab:resultsIHE} we report the maximum observable value of the \BBz half life ($T^{0\nu}_{1/2\,D}$) and the minimum observable value of the Majorana neutrino mass ($|m_{ee}|_D$), according to the given definition of discovery potential, of the IHE. The subscript $D$ stands for discovery and the two values of $|m_{ee}|$ correspond to the most and the least favorable choice of the nuclear factor of merit. 
For the sake of completeness and to ease the comparison with future experiments that quote sensitivity instead of discovery potential, we report in Table~\ref{tab:resultsIHE} also the expected sensitivity in 5 years ($T^{0\nu}_{1/2\,S}$) and the corresponding range of Majorana neutrino mass ($|m_{ee}|_S$). The subscript $S$ stands for sensitivity. The sensitivity is calculated as follows:
\begin{itemize}
\item for \zincato, using the 1.64$\sigma$ (90\% C.L.) background fluctuation Poisson method described in Ref.~\cite{CUOREsensi} 
\item for the remaining isotopes, using the zero-background approximation of Eq.~\ref{eq:sensitivityzero}. This is because the expected number of background counts in the ROI is well below one, so the zero-background condition applies. 
\end{itemize}

\begin{table*} 
\caption{Maximum observable value of the \BBz half life ($T^{0\nu}_{1/2\,D}$) and minimum observable value of the Majorana neutrino mass ($|m_{ee}|_D$) of the IHE, defined as a CUORE-like (988 \ciccio crystals) double beta experiment with bolometers based on the four \BBz candidates discussed in this work. The assumptions are: 90\% enrichment in the candidate isotope; $\Delta E$= 5 keV; 5 y exposure time; background equal to the maximum value between 0.1 cnts/ton/y and the 2$\nu\beta\beta$ pile-up-induced background for $\Delta T $= 1 ms (Table~\ref{tab:pile-up}). The two values of $|m_{ee}|_D$ correspond to the most and the least favorable choice of the nuclear factor of merit. The last two columns reports the 1.64$\sigma$ (90\% C.L.) sensitivity in 5 years (see text for an explanation on how the sensitivity is calculated) and the corresponding range of Majorana neutrino mass. We have added a subscript $S$ to distinguish the values of the Majorana neutrino masses calculated from sensitivity ($|m_{ee}|_S$) and those derived from discovery potential ($|m_{ee}|_D$). }
\begin{center}
\begin{tabular}{ccccccc}
\hline\hline
\\
Crystal 		& IHE mass 	& Exposure  &  $T^{0\nu}_{1/2\,D}$ & $|m_{ee}|_D$	 & $T^{0\nu}_{1/2\,S}$  & $|m_{ee}|_S$ \\
 	 		& [ton]      	         &      [ton$\cdot$y] & [10$^{27}$y]  & [meV]         	 &         [10$^{27}$y]   & [meV]  \\
\\
\hline
\\
ZnSe  		& 0.664		&	    3.3       	  &  0.81 & 18 - 52		& 2.2    & 9 - 27 \\
\tungstato	 	& 0.985		&	    4.9	  &  0.49 & 24 - 45		& 1.5    & 12 - 22 \\
\zincato 		& 0.540		&	    2.7           &  0.19 & 24 - 69		&  0.65 & 11 -  31\\
\teod  		& 0.751		&	    3.7           &  0.90 & 17 - 43 		& 2.6    &  8 - 21 \\
\\
\hline\hline
\end{tabular}
\end{center}
\label{tab:resultsIHE}
\end{table*}

To explore the potential of future experiments, we can speculate on the possibility of completely eliminating the background coming from radioactivity and leaving only the intrinsic contribution of \BBdn. Given the high energy resolution of bolometers, the background from the tail of the \BBd decay is negligible. The ultimate contribution would then be the \BBd pile-up for $\Delta T $= 1 ms whose rate expressed in cnts/ton/y in a 5 keV energy window is listed in Table~\ref{tab:pile-up}.  

\begin{figure*}
\begin{center}
\includegraphics[width=0.5\textwidth,clip=true]{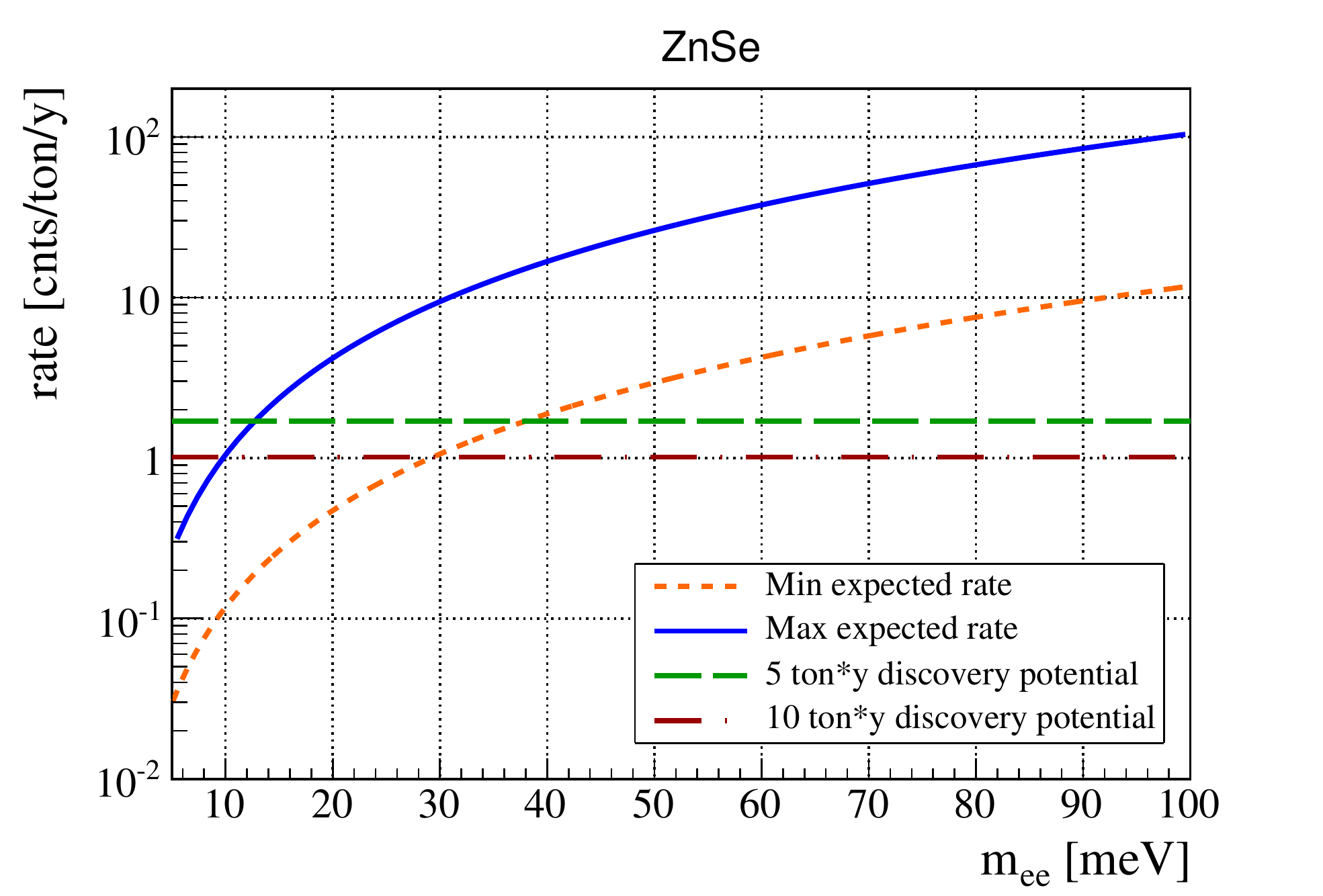}
\includegraphics[width=0.5\textwidth,clip=true]{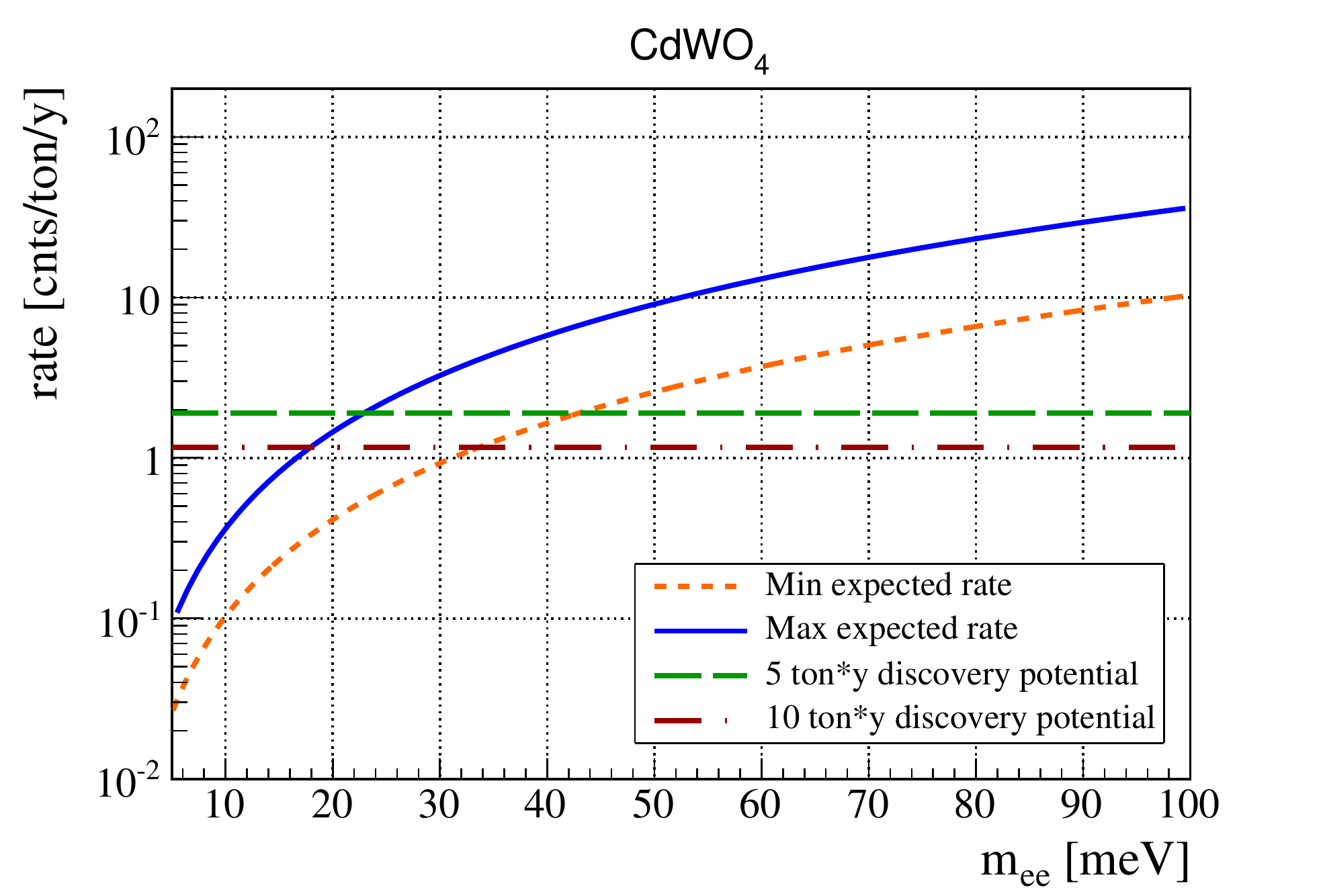}
\includegraphics[width=0.5\textwidth,clip=true]{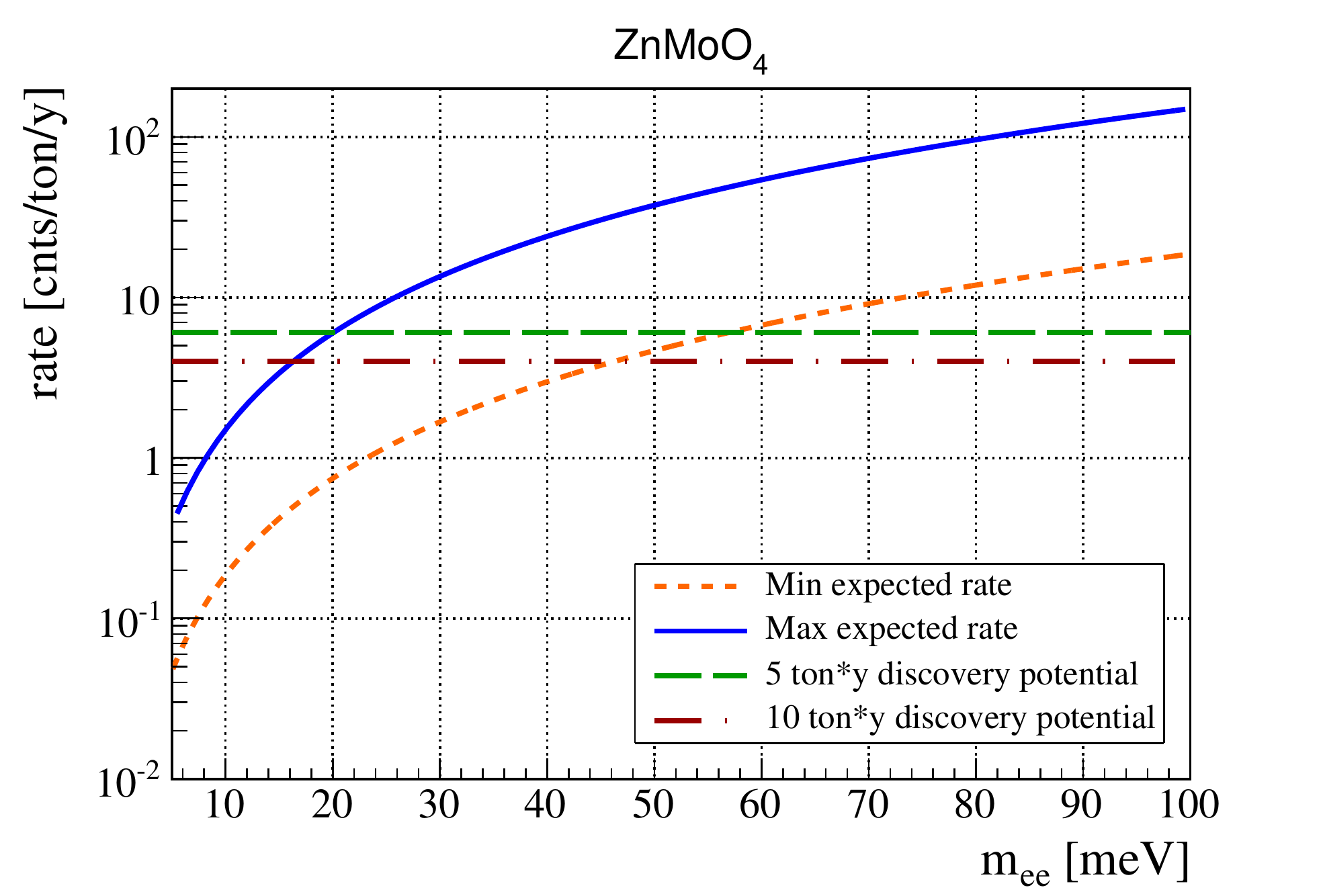}
\includegraphics[width=0.5\textwidth,clip=true]{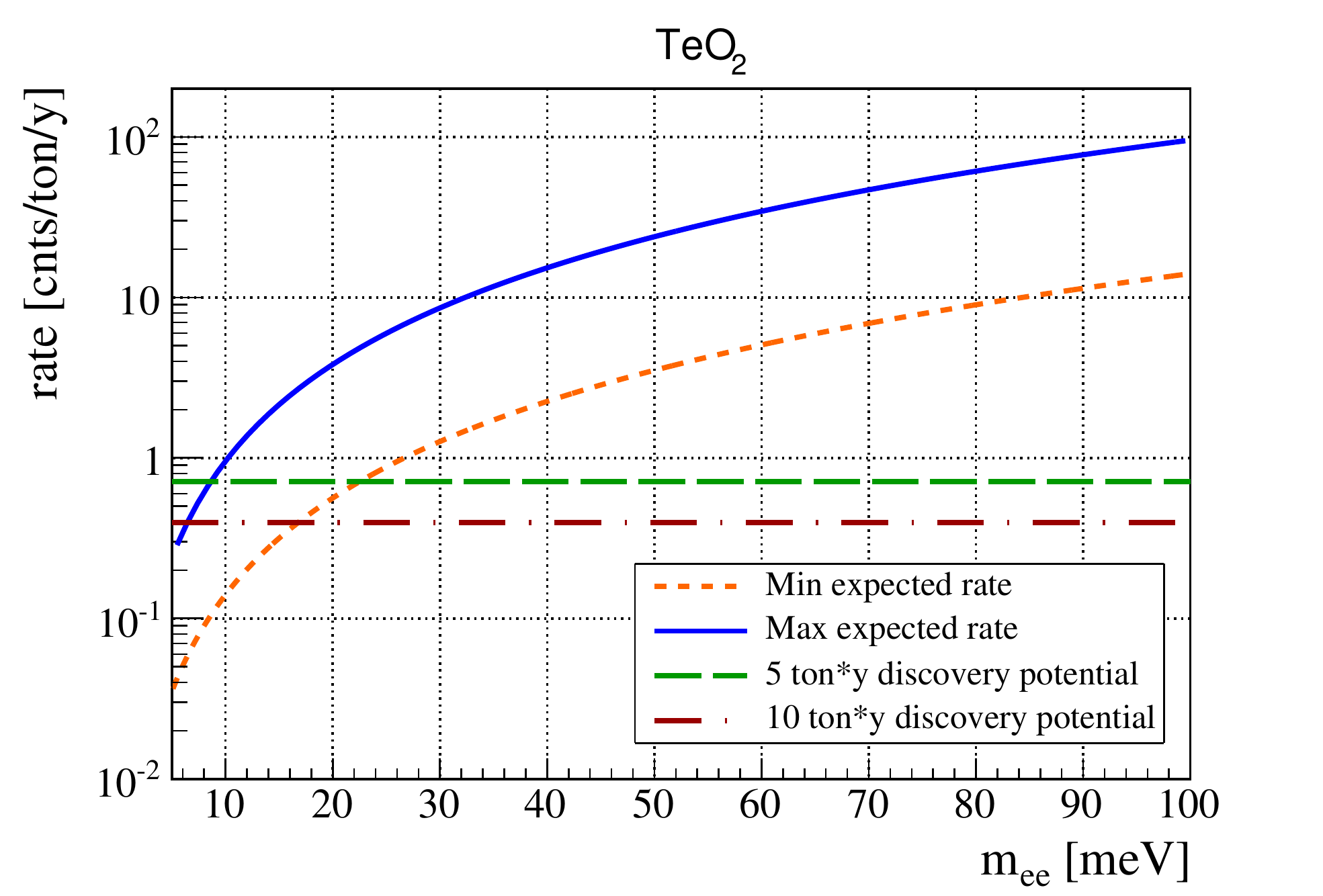}
\end{center}
\caption{Discovery potential (see text) of future double beta experiments with bolometers based on the four \BBz candidates discussed in this work. We assume that the only contribution to the background comes from the irreducible \BBd pile-up.  A summary of these results is given in Table~\ref{tab:results}.}
\label{fig:results}
\end{figure*}

In Fig.~\ref{fig:results} we plot for each candidate isotope, the predicted \BBz rate vs. the Majorana neutrino mass ($|m_{ee}|$) for the most favorable (blue) and least favorable (orange) values of $F_{N}$ presented in Fig.~\ref{fig:FN}. The green and red horizontal lines represent the observable rate defined according to our discovery criteria, for an exposure of 5 ton$\cdot$y and 10 ton$\cdot$y respectively. The values of $|m_{ee}|$ where each horizontal line intersects the curves are the minimum observable value (for the given exposure) of the Majorana neutrino mass for the most and the least favorable choice of the nuclear factor of merit. 

A summary of the results presented in the plots of Fig.~\ref{fig:results} is given in Table~\ref{tab:results}.  For each candidate isotope we report the maximum observable value of the \BBz half life ($T^{0\nu}_{1/2\,D}$) and the minimum observable Majorana mass ($|m_{ee}|_D$) for the most and least optimistic choice of the nuclear matrix elements for the cases of both a 5 ton$\cdot$y and 10 ton$\cdot$y exposure. As we did for Table~\ref{tab:resultsIHE} we report in Table~\ref{tab:results} also the expected sensitivity ($T^{0\nu}_{1/2\,S}$) and the corresponding range of Majorana neutrino mass. We have added a subscript $S$ to distinguish the values of the Majorana neutrino masses calculated from sensitivity ($|m_{ee}|_S$) and those derived from discovery potential ($|m_{ee}|_D$).

\begin{table*}
\caption{Maximum observable value of the \BBz half life ($T^{0\nu}_{1/2\,D}$) and minimum observable value of the Majorana neutrino mass ($|m_{ee}|_D$) of a future double beta experiment with bolometers based on the four \BBz candidates discussed in this work. The assumptions are: 90\% enrichment in the candidate isotope; $\Delta E$= 5 keV; 5 y or 10 y exposure time; background equal to the 2$\nu\beta\beta$ pile-up-induced background for $\Delta T $= 1 ms (Table~\ref{tab:pile-up}). The two values of $|m_{ee}|$ correspond to the most and the least favorable choice of the nuclear factor of merit. TThe last two columns reports the 1.64$\sigma$ (90\% C.L.) sensitivity in 5 years (see text for an explanation on how the sensitivity is calculated) and the corresponding range of Majorana neutrino mass. We have added a subscript $S$ to distinguish the values of the Majorana neutrino masses calculated from sensitivity ($|m_{ee}|_S$) and those derived from discovery potential ($|m_{ee}|_D$).}
\begin{center}
\begin{tabular}{cccccc}
\hline\hline
\\
Crystal 					& Exposure 	&  $T^{0\nu}_{1/2\,D}$ & $|m_{ee}|_D$	&  $T^{0\nu}_{1/2\,S}$	&	$|m_{ee}|_S$	 \\
 	 					& [ton$\cdot$y]      	& [10$^{27}$y]  & [meV]         	&       [10$^{27}$y] 	& [meV]	 \\
\\
\hline
\\
\multirow{2}{*}{ZnSe}  		& 5 		& 1.5 	& 13 - 38		& 3.3 & 8 - 22 \\
 						& 10		& 2.6	 	& 10 - 29		& 6.5	 & 5 - 15 \\
\multirow{2}{*}{\tungstato}	 	& 5		& 0.55	& 23 - 43		& 1.5 & 12 - 22 \\
					 	& 10		& 0.89	& 18 - 34		& 3.0 & 8 - 15 \\
\multirow{2}{*}{\zincato} 		& 5		& 0.27 	& 20 - 57		& 0.9 & 9 - 27 \\
					 	& 10		& 0.41 	& 16 - 46		& 1.4 & 8 - 21 \\
\multirow{2}{*}{\teod}  		         & 5		& 3.3 	& 9 - 23 		& 3.4 & 7 - 18 \\
						& 10		& 5.9 	& 6 - 17 		& 6.8 & 5 - 13 \\
\\
\hline\hline
\end{tabular}
\end{center}
\label{tab:results}
\end{table*}

Considering the variations due to different definitions of the discovery potential (there is always a certain degree of freedom in this choice, as discussed in~\cite{CUOREsensi}) we can conclude that the experimental technique based on $\alpha$-discrimina\-ting bolometers grown from enriched material may lead to the investigation of a large fraction of the IH region. Some cases, such as \teod with Cherenkov light readout or ZnSe with 10 years exposure, may lead to complete IH exploration.  However, given the strong impact of NME uncertainties on $|m_{ee}|$ limits, future developments in the calculations of NME might change the situation considerably. 

Experimental goals discussed above require tremendous effort towards background reduction.  Of particular concern is the understanding and control of surface contamination on the various detector materials.  To contribute to this effort, the feasibility of high efficiency $\alpha$-discrimination in \teod bolometers using Cherenkov light readout should be demonstrated;  if possible, both Cherenkov and scintillating bolometer techniques should be scaled to a large size experiment.  Light detector performance, such as energy- and time-resolution as well as light collection efficiency, should also be improved. All these items are already the subject of an intense R\&D and new results will be available soon.  Lastly, economics of the isotopic enrichment, feasible for all the candidate isotopes described in this work, needs to be understood.
%
\section*{Conclusions}
%
Currently operating and near-term \BBz experiments will not be able to explore the large portion of the IH region. Limiting factors are the relatively small isotope masses and/or the large background counting rates. Future plans in this field will require the use of high resolution detectors to deal with the intrinsic background induced by \BBd decay.

In this paper, we have studied the possibility of a particular type of bolometer: those that are able to discriminate $\alpha$-induced backgrounds from $\beta/\gamma$ backgrounds. These high resolution devices not only efficiently reject \BBd background, but their $\alpha$-rejection capability also makes them mu\-ch less sensitive to the surface contamination that presently dominates the background in experiments based on cryogenic bolometers like CUORE. 

Assuming that the sensitivity is limited only by the undiscriminated pile-up of \BBd events, we have demonstrated that the use of $\alpha$-discriminating bolometers, grown from enriched material, can lead to a feasible future experiment having a \BBz discovery potential in the IH region.  We have also addressed areas where the strongest effort will be required in order to improve the performances of such detectors. We conclude that complete coverage of the IH region is possible with a large array of low-background bolometric detectors. Achieving this goal requires significant isotopic mass of order of 1 ton, and nearly zero background rates in the \BBz region of interest. Active background rejection techniques, such as pulse shape discrimination or detection of scintillation or Cherenkov light, offer a promising way to suppress the most dominant $\alpha$ backgrounds. The ultimate sensitivity of such detectors may be limited by the accidental pile-up between multiple \BBd events in the same crystal. Capability of rejecting such pile-up events in the bolometric detectors needs to be demonstrated.

%
\section*{Acknowledgments}
The CUORE Collaboration thanks the directors and staff of the Laboratori Nazionali del Gran Sasso and the technical staff of our laboratories. This work was supported by
the Istituto Nazionale di Fisica Nucleare (INFN); the Director, Office of Science, of the U.S. Department of Energy under Contract Nos. DE-AC02-05CH11231 and DE-AC52-07NA27344; the DOE Office of Nuclear Physics under Contract Nos. DE-FG02-08ER41551 and DE-FG03-00ER41138; the National Science Foundation under Grant Nos. NSF-PHY-0605119, NSF-PHY-0500337, NSF-PHY-0855314, \\NSF-PHY-0902171, and NSF-PHY-0969852; the Alfred P. Sloan Foundation; the University of Wisconsin Foundation; and Yale University. This work was partially supported by ERC (FP7/2007–2013) under project Lucifer, grant agreement n. 247115.
%

%
\section*{Bibliography}
%
\bibliographystyle{apsrev}

\end{document}